\documentclass[aps.pra,twocolumn,amsmath,amssymb,showpacs,showkeys,floatfix]{revtex4-2}
\usepackage{graphicx}
\usepackage{amsfonts}
\usepackage{epsfig}
\usepackage{dcolumn}
\usepackage{amsthm}
\usepackage{color}
\usepackage{physics}
\usepackage{latexsym}
\usepackage{amscd}
\usepackage{relsize}
\usepackage{import}
\makeatletter
\newcommand*{\balancecolsandclearpage}{%
  \close@column@grid
  \cleardoublepage
}

\begin{document}


\title{Extreme Events of Quantum Walks on Graphs}
\author{Nisarg Vyas $^1$}
\email{nisarg.vyas@students.iiserpune.ac.in}
\author{M. S. Santhanam $^1$}  
\email{santh@iiserpune.ac.in}
\affiliation{
$^1$ Physics Department, Indian Institute of Science Education and Research, Pune
}

\date{\today}

\begin{abstract}
Due to the unitary evolution, quantum walks display different dynamical features from that of classical random walks. In contrast to this expectation, in this work, we show that extreme events can arise in unitary dynamics and its properties are qualitatively similar to that of random walks. We consider quantum walks on a ring lattice and a scale-free graph. Firstly, we obtain quantum version of flux-fluctuation relation and use this to define to extreme events on vertices of a graph as exceedences above the mean flux. The occurrence probability for extreme events on scale-free graphs displays a power-law with the degree of vertices, in qualitative agreement with corresponding classical random walk result. For both classical and quantum walks, the extreme event probability is larger for small degree nodes compared to hubs on the graph. Further, it is shown that extreme event probability scales with threshold used to define extreme events. 

\end{abstract}

\maketitle

Quantum walks (QW) were introduced as the quantum counterpart to the classical random walk (CRW) models \cite{qrw_aharonov}. In practice, quantum walks represent unitary evolution applied to an initial wave-packet and is strongly reliant on the peculiar features of quantum mechanics -- superposition, entanglement and even measurements. Thus, quantum walks are distinct from classical random walks. For instance, a basic feature of the standard CRW on a line is that the ``distance'' covered by a walker in $N$ time-steps is $\sim \sqrt{N}$, while for a quantum walker it is $\sim N$. In the context of quantum algorithms, this property provides a basic template for quantum induced speed-ups reported in the literature \cite{Ambainis_quant_algo} for tasks ranging from search and transport on networks \cite{childs2003_expoHit, Kempe2002_expoHit, Farhi1997_expoHit, M_lken_2006}, element distinctness \cite{ambainis2014_distinctness} and triangle-finding problems \cite{magniez2005_triangleFind}. Furthermore, quantum walk proposals have been experimentally realized using single photon dynamics to integrated photonics \cite{YanZhaGon2019,FlaSpaSci2018, BroFedLan2010,SchCasPot2011,SanSciVal2012,CreOseRam2013}, and cold atomic traps \cite{MicLeoJai2009,ZahKirGer2010,PreMaTai2015}. 

Despite these impressive theoretical and experimental developments, a possibility of emergent phenomena in quantum walks had not received any attention with one exception being the localization induced by the presence of some form of disorder \cite{DudIvaSah2023,SchCasPot2011,ShaBoe2022,CreOseRam2013,LiIzaWan2013,YaoWald2023,ZengYong2017}. The classical random walk with multiple random walkers display a plethora of emergent phenomena \cite{Sethna_emergent,FiGaretto2017,SturhmannCoghi2024, LechRobMann2024}, among which extreme events (EE), {\it i.e.,} significant deviations away from the mean behavior, is of interest in this paper. In the classical setting of multiple random walkers on random networks, the dynamics of extreme events is of practical interest in the context of  phenomena such as traffic jams, power blackouts or market crashes \cite{BarloSchadShreck2001, GandhiSanthanam2022, KishoreSanthanamAmritkar2012,AanjSanthanamKulkarni2020}. What would be the equivalent extremes in the quantum setting ? If the probability of quantum walker at each site indicates the magnitude of ``event'', then what is the probability that extreme events -- events larger than certain pre-determined size -- takes place? Such questions have begun to attract attention only in the recent times. For example, EE was shown to arise in quantum walks performed on a discrete lattice in which phases of the walker's amplitudes are randomized at every time step \cite{rougewave_1d,BuaDiaAlm2023,BuaRap2023}. The rogue waves can be described in terms of Gumbel distribution, one of the classical extreme value distributions. Further, it was argued that phase-disorder induced quantum walks can be controlled by tuning the coin operator \cite{BrunCartretAmbainis2003,BrunCartretAmbainis2003_QtoCl_transition, Arul2021_quantumChaoticCoins}.

Introducing phase disorder in quantum walks brings it closer to classical regime since the delicate phase relations between the amplitudes are weakened \cite{Kendon2007}. Hence, the emergence of large amplitudes cannot necessarily be attributed to coherent effects of the quantum walk process. Then, a natural question arises if quantum walks can inherently exhibit large amplitudes without requiring introduction of randomness in the phases. Such large amplitude waves in quantum systems had been reported in the dynamics of Bose-Einstein condensates governed by nonlinear Schrodinger equation \cite{BluKonAkh2009, YanKonAkh2010, DemGraOno2019, DinZhoMal2024, KunGhoRoy2022}. In quantum transport on one-dimensional lattices, governed by the linear Schrodinger equation, uncorrelated disorder at lattice sites induces Anderson localization, while correlated disorder leads to mobility edge \cite{IzrKro1999}. In general, the relative roles of randomness and non-linearity at generating EEs has been a long-standing debate \cite{santhanam, deterministic_nonLin_EE, EE_inhomogeneity}. In this work, we focus on large amplitudes of discrete time quantum walks (DTQW) generated by linear unitary evolution. We show that, contrary to expectations, the extreme event statistics of amplitudes qualitative follows that of classical random walks on networks.

\begin{figure*}[th]
    \includegraphics*[width=1.0\linewidth]{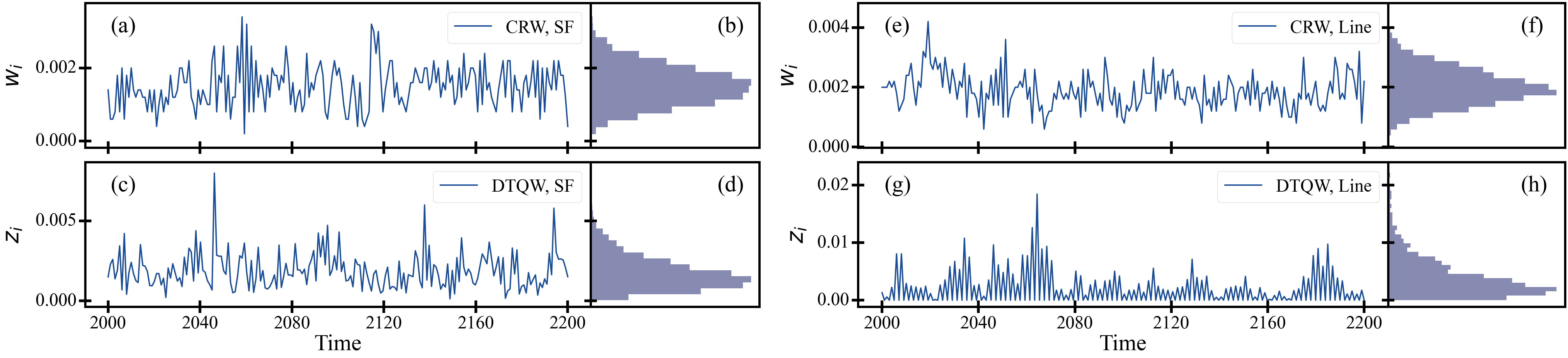}
\caption{Time-series of  the normalized occupation probabilities ($w_i/W$) and squared probability amplitudes ($z_i$) for CRW and DTQW respectively. Figures (a)-(b) show that for a node of degree 3
on a scale-free graph of size 501, while figure (c)-(d) are for a periodic 1D lattice of size 501.}
\label{fig:spread}
\end{figure*}

Discrete-time quantum walk (DTQW) is a unitary operation $U = SC$ applied on the initial state $\psi(t=0)$ representing the walker, where $C$ and $S$ represent a coin and a shift operation \cite{aharonov}. The state of the walker at time $t>0$ is given by $\psi(t) = U^t \psi(0)$. On a regular graph with $k$ edges at each vertex, $S$ and $C$ are independent of the vertex, while on a general disordered and undirected graph (such as a scale-free graph) the model requires a vertex-dependent coin and shift operators (the operators shown in SI). The choice of the coin and the shift operator determines properties of quantum walk on a given graph \cite{godsil}. In this work, we obtain results for QW on a scale-free (SF) graph with degree distribution $P(k) \sim k^{-2.3}$, and on a ring graph. 

Since the central object of interest is extreme events, firstly it is instructive to visualize the ``events''. Let us consider SF graph with $N=500$ vertices and degree distributed as $P(k) \sim k^{-2.3}$. On this graph, $W$ walkers execute independent CRW dynamics and the ``events'' are the instantaneous number of walkers $w_i(t)$ on $i$-th vertex. In Fig. \ref{fig:spread}(a), as would be anticipated, $w_0(t)/W$ at vertex labeled $0$ is seen to display stochastic fluctuations. In all simulations reported here, all the $W$ walkers at time $t=0$ are initially placed at one node and evolution up to $t=2000$ are ignored as transients. Figure \ref{fig:spread}(b) shows the distribution of walkers, and it is known to be a Binomially distributed \cite{santhanam}. Next, DTQW is executed on the same graph with initial state $\psi(0) = \delta(0)$. In this quantum case, the ``events'' are the probability density $z_i(t) = |\psi_i(t)|^2$  at $i$-th vertex. As observed in Fig. \ref{fig:spread}(c), it has the characteristics of a stochastic process though the underlying dynamics is unitary and deterministic. The corresponding distribution is displayed in Fig. \ref{fig:spread}(d). Note that even this cursory visual examination shows that the width of the distribution is larger for the DTQW compared to that of the CRW.
Now, we examine the events for walks on a periodic ring with $N=500$ vertices and degree of each node being $k_i=2, i=1,2, \dots N$. Figure \ref{fig:spread}(e) shows $w_0(t)/W$ for CRW, and the anticipated stochastic fluctuations are observed, and its distribution is depicted in Fig. \ref{fig:spread}(f). For DTQW on the same ring graph (Fig. \ref{fig:spread}(g)), we observe a similar but possibly more correlated fluctuations. The corresponding distribution is shown in  Fig. \ref{fig:spread}(h). This figure is representative of the dynamics observed at all the vertices.

\begin{table}[b]
\caption{Mean and standard deviation deviation for quantum and classical walks on periodic lattice. The sizes of the 1D, 2D and 3D lattice is $N=729$.}
\begin{tabular*}{\linewidth}{@{\extracolsep{\fill}} l ccc}
\hline
\hline
\textbf{Periodic lattice} & \textbf{1D} & \textbf{2D} & \textbf{3D} \\
\hline
\multicolumn{4}{c}{mean flux  ($\times 10^{-3}$)  } \\
\hline
$\langle w \rangle$ (classical) & 1.372 & 1.372 & 1.372 \\ 
$\langle z \rangle$ (quantum) & $1.372$ & $1.372$ & $1.372$ \\
\hline
\multicolumn{4}{c}{Standard deviation of the flux ($\times 10^{-2}$) }\\
\hline
$\sigma_{w}$ (classical)& 3.701 & 3.701 & 3.701 \\ 
$\sigma_{z}$ (quantum) & 0.149 & 0.078 & 0.061 \\
\hline
\end{tabular*}%
\label{tab:1}
\end{table}

Before examining the extreme events, the fluctuation characteristics, namely, the relation between mean flux and standard deviation of the flux passing through $i$-th vertex merit our attention. For CRW, this is an exactly solvable problem. If $w_i$ is the flux through $i$-th vertex, 
and $\langle w_i \rangle$ and $\sigma_i$ represent the mean and standard deviation respectively, and $\langle . \rangle$ represents average over time, then in the limit that $k_i << 2E=\sum_{i=1}^N k_i$, we obtain
\begin{equation}
    \sigma_i = \sqrt{\langle w_i \rangle},    ~~~~\text{where } \langle w_i\rangle = \frac{Wk_i}{2E}.
    \label{eq:ffrel}
\end{equation}
In this, $E=(1/2)\sum_i k_i$ is the number of edges on the graph.
This result in Eq. \ref{eq:ffrel}, called the flux-fluctuation relation, is satisfied by CRW on any graph. 

\begin{figure}
        \centering
    \includegraphics[width=1\linewidth, height=0.5\linewidth]{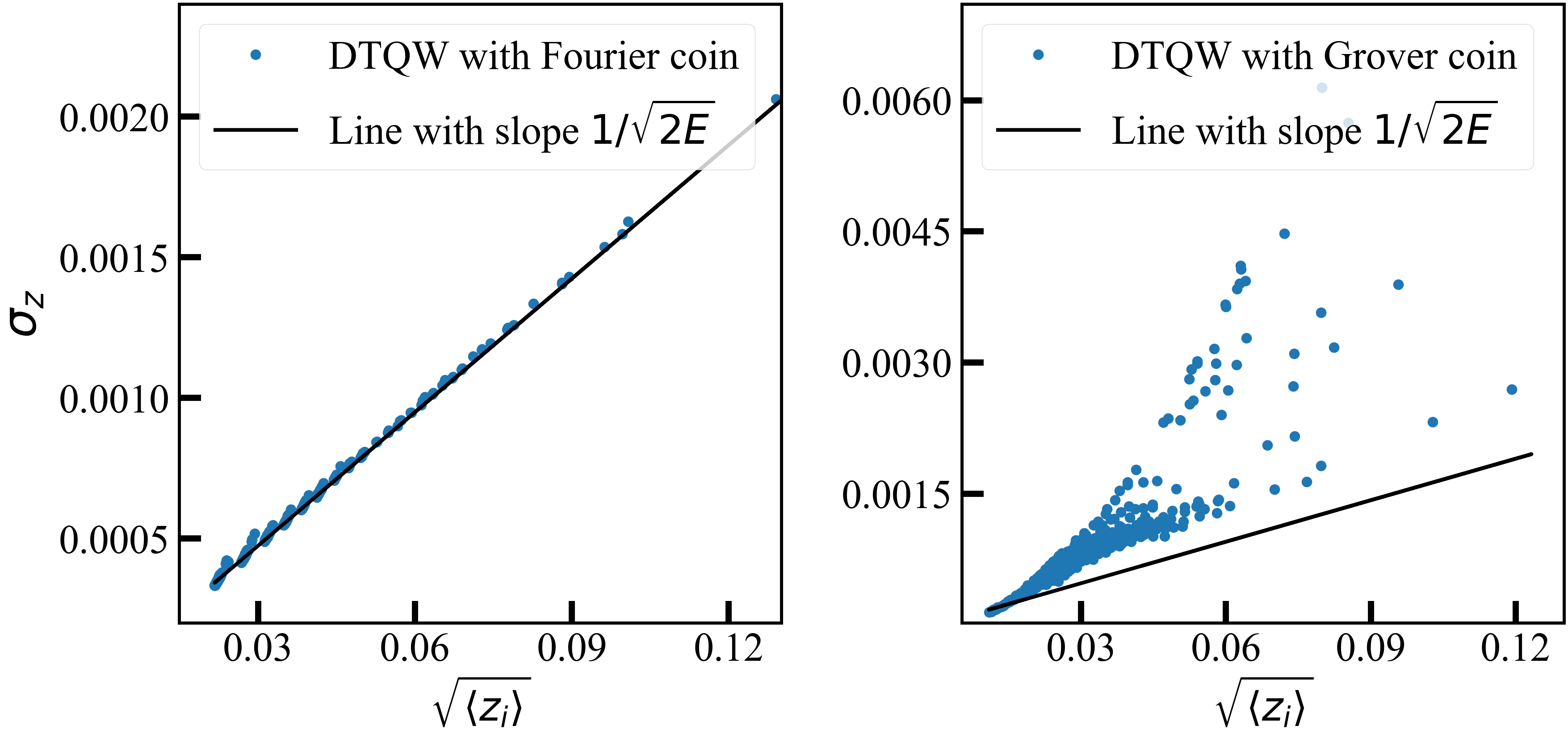} 
    \caption{Mean flux $\langle z \rangle$ vs. standard deviation $\sigma_z$ for DTQW on a scale-free graph with $N=1000$ vertices (circles); (a) DTQW with Fourier coin, (b) DTQW using Grover coin. Solid line indicates expected slope $1/\sqrt{2E}$ based on Eq. \ref{eq:qffrel}.}
    \label{fig:ffrel}
\end{figure}

Does flux generated by DTQW follow this relation? To answer this question, we define mean flux to be $\langle z \rangle$ and standard deviation to be $\sigma_z=\langle (z-\langle z \rangle)^2 \rangle$, where $\langle . \rangle$ represents time average. As the quantum walk is a unitary evolution, we expect it to strongly deviate from Eq. \ref{eq:ffrel}. As observed in Table \ref{tab:1} for a DTQW with periodic boundary condition on regular lattices in 1- to 3-dimensions, $\langle z \rangle$ is nearly identical to that of CRW, though $\sigma_z$ deviates from the classical result. Figure \ref{fig:ffrel} shows the flux-fluctuation relation for DTQW on a SF graph. Remarkably, $\langle z \rangle$ and $\sigma_z$ have a linear relation, in qualitative agreement with the case of CRW. However, the slopes are different; it is 1 for the CRW, and $1/\sqrt{2E}$ for the DTQW. This gives the first hint that some aspects of quantum walk, at least qualitatively, have a similarity with classical relations. In rest of this paper, we explore the consequences of this for extreme events and also show why coherent quantum phenomenon should have qualitative resemblance to the classical relations such as in Eq. \ref{eq:ffrel}.

Now, we obtain the flux-fluctuation relation analytically for DTQW. To perform a DTQW on an arbitrary graph $G(V,E)$ with a vertex set $V$ and an edge set $E$, the undirected graph is viewed as a directed one and the quantum walk takes place in a complex vector space $\mathbf{C}^{2\vert E \vert}$ spanned by the basis $e_{ij}$ representing the edge between vertices $i$ and $j$. The shift operator $\widehat{S}$ is defined through its action on the basis elements
\begin{equation}
    \widehat{S} ~ e_{i,j} = e_{j,i},
    \label{eq:shift_def}
\end{equation}
so as to swap the amplitudes on the edges. The coin at each vertex acts to replace the amplitude on each edge with a superposition of that on all the other edges of $u$. For a graph with $N$ vertices, $\vert V \vert  = N$, and the coin operator $C$ is
\begin{equation}
C = {\rm diag}\{ C_1, C_2, ..., C_N \}    ,
\end{equation}
where each $C_i$ is a unitary matrix of order $k_i$, degree of $i$-th vertex. In this work, a model of DTQW is used which fixes the shift operator, and allows a choice of coin operator at each vertex \cite{godsil}. However, our analytical results depend solely on the spectral properties of $U$ and hence are not limited by this choice of the shift operator. For the numerical results we use the shift operator in Eq. \ref{eq:shift_def} in combination with Fourier and Grover coins to demonstrate the dependence on the coin spectra. 

Unitary evolution of $U$ implies that an initial state vector $\lvert x \rangle$ remains normalized and does not converge to any stationary distribution unless $U\lvert x \rangle = \lvert x \rangle$. The walker probability amplitude at vertex $v$ is the sum of the probabilities on the edges incident on $v$. If the initial state is $\lvert x \rangle$, the probability amplitude on $v$ at time $t$ is
\begin{equation}
\label{eq:prob_timeseries}
    z_{v}(t) = \langle x \rvert \left(U^t\right)^* D_v U^t \lvert x \rangle,
\end{equation}
where $D_v$ is the diagonal matrix such that $D_{ii}=1$ if the $i^{th}$ entry of the state vector represents some outgoing edge of $v$. That is, $D_v$ projects the amplitudes on a subspace spanned by the edges of $v$. The probability in Eq. \ref{eq:prob_timeseries} is time dependent, though the time-averaged probability distributions
\begin{equation}\label{eq:convergence}
    \lim_{T\to\infty} \frac{1}{T} \sum_{t=0}^{T-1} z_{v}(t) = \sum_r \langle x \rvert F_r D_v F_r \lvert x \rangle
\end{equation}
can converge \cite{aharonov}. In this, $F_r$ are the spectral idempotent of $U$, and $U$ is assumed to have a non-degenerate eigen-spectrum. Irrespective of whether the initial state is chosen to be localized on one vertex or uniformly spread out over all of the vertices, the time-averaged probability amplitudes, averaged further over vertices of same degree converges in the limit of $|V|\gg1$. Detailed derivation is presented in SI \cite{supp_material}, and the final result is
\begin{equation}
\langle z_{v} \rangle = \frac{1}{\vert S_k\vert}\sum_{\substack{r \\ v\in S_k}} \langle x \rvert \left(F_r D_{v} F_r\right) \lvert x \rangle = \frac{k}{2E} + \mathcal{O} \left(\frac{1}{E^2} \right),
\label{eq:tavgprob}    
\end{equation}
where $S_k = \{v\in V \vert deg(v) = k\}$. This equation states that the mean probability density (at a vertex with degree $k$) is proportional to degree of the vertex being considered. This is qualitatively similar to Eq. \ref{eq:ffrel} for CRW.  Further, assuming eigenvalues of $U$ to be non-degenerate and if the initial state is a uniform superposition over all the edges, we get standard deviation $\sigma_z$ as (details of the calculation given in SI)
\begin{equation}
\sigma_z \sim \frac{\sqrt{k}}{2E}.    
\label{eq:sigmaz}
\end{equation}
Quite remarkably, combining Eq. \ref{eq:tavgprob} and \ref{eq:sigmaz}, we get a quantum equivalent of flux-fluctuation relation:
\begin{equation}
\sigma_z \approx \frac{\sqrt{\langle z_{v} \rangle}}{\sqrt{2E}}.
\label{eq:qffrel}
\end{equation}
This is qualitatively similar to the classical relation in Eq. \ref{eq:ffrel}, and is verified through the simulation shown in Fig. \ref{fig:ffrel}(a). The straight line in Fig. \ref{fig:ffrel}(a) showing the relation between $\sigma_z$ and $\langle z_{v} \rangle$, has slope $1.60 \times 10^{-2}$, consistent with the expected slope of $1/\sqrt{2E} \sim 1.50 \times 10^{-2}$. 

What happens if the spectra of $U$ is degenerate? From Eq. \ref{eq:prob_timeseries}, it must be noted that
\begin{equation}
\left( U^t \right)^* D_v U^t = \sum_r F_r D_v F_r + \sum_{r \neq s} e^{i t\left(\theta_s-\theta_r\right)} F_r D_v F_s.
\end{equation}
The result in Eq.\ref{eq:tavgprob}-\ref{eq:sigmaz} were derived from the first term since non-degeneracy of eigenspectrum of $U$ guarantees that the second term has a vanishing contribution (see SM). However in the case of degenerate eigenvalues, the the second term also contributes to the time-averaged $\langle z \rangle$ at a vertex. Then, the DTQW strongly deviates from the classical walk. Both $\langle z_v \rangle$ and $\sigma_z$ significantly differ from corresponding classical values, and the flux-fluctuation relation is no more linear. This can be seen in the case of DTQW with Grover coin (in Fig. \ref{fig:ffrel}(b)) which has large degeneracy for the eigenvalues $\pm 1$, and consequently shows poor agreement with both the classical results and quantum flux-fluctuation relation (\ref{eq:qffrel}).

\begin{table}[t]
\caption{Extreme event probabilities for quantum and classical walks on periodic lattices. The sizes of the 1D, 2D and 3D lattice is 729. The total number of walkers for the CRW is 100, while the thresholds for EE was set at $m=3$.}
\begin{tabular*}{\linewidth}{@{\extracolsep{\fill}} l ccc} 
\hline
\hline
\textbf{Periodic lattice} & \textbf{1D} & \textbf{2D} & \textbf{3D} \\
\hline
$\langle \mathcal{F}_i \rangle$ (classical) & $0.1283$ & $0.1283$ & $0.1283$ \\ 
$\langle \mathcal{F}^q_i \rangle$ (quantum) & $0.0202$ & $0.0135$ & $0.0109$ \\ \hline
\end{tabular*}
\label{tab:2}
\end{table}
Next, we focus on the occurrence of extreme events. Consider this classical scenario: $W$ walkers on a finite graph with $N$ vertices and $E$ edges are executing CRW. 
If $w_i(t)$ represent the magnitude of ``event'' on $i$-th vertex, following Ref. \cite{santhanam}, this event would be designated as extreme event if $w_i(t) > w_i^{\rm ee}$, where $w_i^{\rm ee} = \langle w_i \rangle + m \sigma_i$, with $m \in \mathbb{R}$, represents vertex-dependent threshold for extreme events. Then, probability for the occurrence of extreme events is $F_i^c = {\rm prob}(w_i>w_i^{\rm ee})$. For DTQW,  the probability density $z_i(t) = |\psi_i(t)|^2$ represents magnitude of ``event'' at time $t$. Then, analogous to the classical case, extreme event is based on the condition $z_i(t) > z_i^{\rm ee}$, where the threshold is $z_i^{\rm ee} = \langle z_i \rangle + m \sigma_z$. Corresponding extreme event occurrence probability for DTQW is:
\begin{equation}\label{eq:ee_definition}
    F_i^q = {\text{prob}}(z_i>z_i^{\text{ee}}).
\end{equation}
In principle, $F_i^q$ can be obtained from the distribution of $z$, which does not appear to be analytically tractable. Hence, we estimate $F_i^q$ from finite time series of $z_i(t)$. 

Table \ref{tab:2} shows numerically estimated extreme event probability on regular graph with periodic boundary conditions. For CRW, $k_i=k$, a constant for all the vertices. Then, $\langle w \rangle = W/(N)$ (due to Eq. \ref{eq:ffrel}), that is, the extreme event probability depends only on $W$ and $N$, and not on the dimensionality. Thus, even when the graph dimension increases, $\mathcal{F}^c_i$ remains invariant as seen in Table \ref{tab:2}. For quantum walks, $\langle z \rangle$ equals $\langle w \rangle$ (we refer the reader to \cite{aharonov} for the details), but $\sigma_z \ne \sigma_w$ (see Table \ref{tab:1}), and increases with dimensionality. Hence, $\mathcal{F}^q$ for QW decreases as dimensionality of lattice increases.

\begin{figure}
\centering
\includegraphics[width=0.85\linewidth]{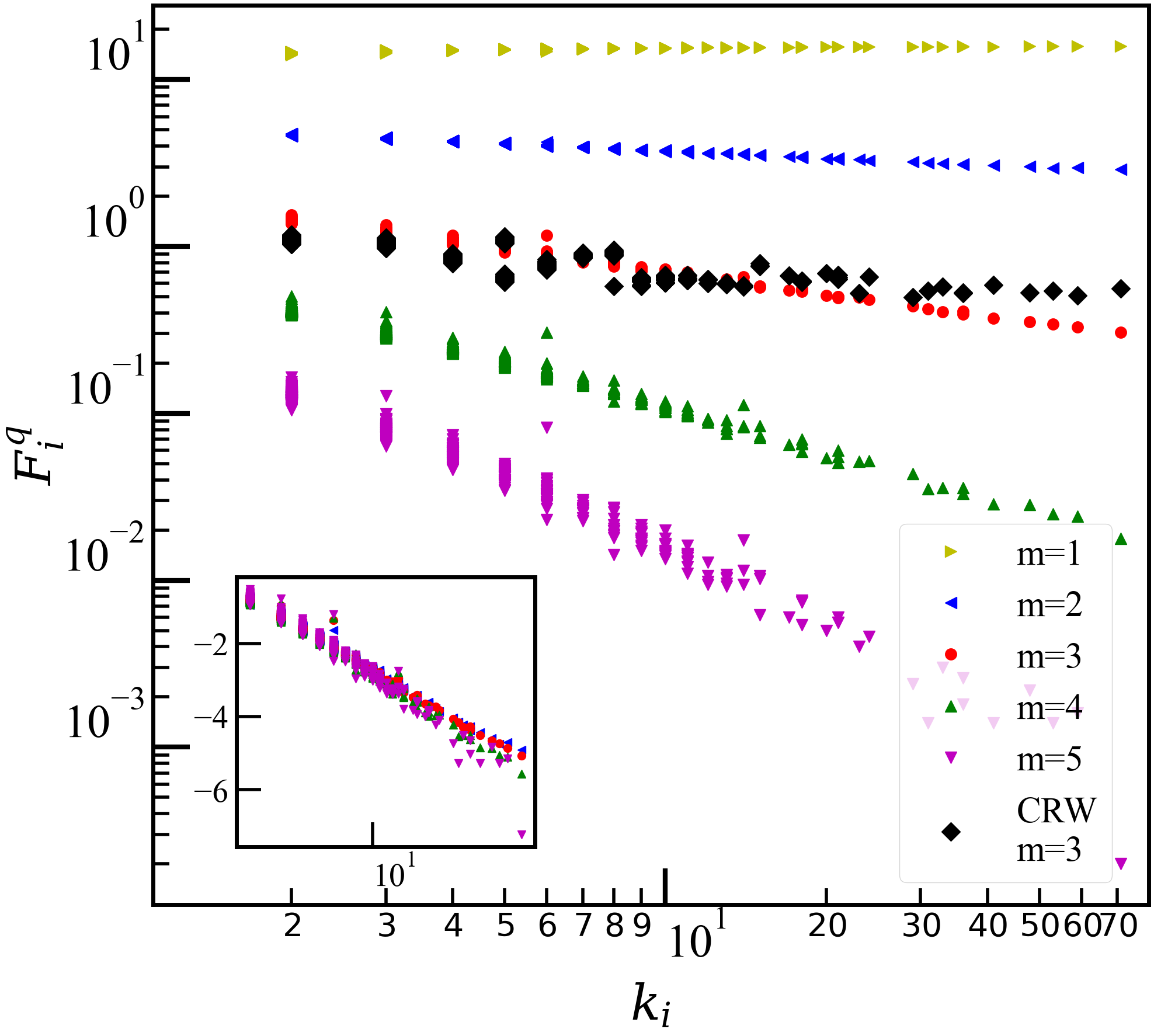}
\caption{Extreme event probabilities as a function of degree $k$ and threshold $m$ (log-log scale) for both DTQW and CRW on SF graph with $N=1000$ vertices. There is exactly one node of degree 71 (highest) while there are 505 nodes of degree 2 (smallest). For CRW, we show only the $m=3$ threshold results for visual clarity, however the results for other thresholds follow a similar trend. (Inset) shows scaled EE probabilities as a function of degree, where the scaling relation was numerically found to be $F_i^q \sim k^{\gamma_m}$, where $\gamma_m = 0.17 - 0.09m$.}\label{fig:eeprob}
\end{figure}

Now, let us examine the extreme event probability on SF graph with $N=1000$. Figure \ref{fig:eeprob} shows $F_i^q$ as a function of degree $k$ for DTQW for several values of $m$, which defines EE threshold. The EE probabilities decay as a power-law with degree $F^q_i \sim k^{\gamma_m}$, with threshold dependent exponent $\gamma_m \approx 0.17 - 0.1 m$. This is reminiscent of a similar trend for the CRW case \cite{santhanam}. Though large degree vertices (hubs) attract more quantum walkers than small degree vertices (note $\langle z \rangle \propto k$), surprisingly, $F_i^q$ for small degree vertices are larger than that for hubs. This is more pronounced when $m > 1$. Figure \ref{fig:eeprob} is also suggestive of scaling of $F_i^q$ with $m$. Indeed, as displayed in the inset of Fig. \ref{fig:eeprob}, we observe a good collapse of all the curves onto one. This implies that the decay trend of extreme event probability is independent of the choice of threshold.

The results in Figs. \ref{fig:ffrel}-\ref{fig:eeprob}, taken together, suggest that as far as extreme events are concerned, quantum walk results are qualitatively similar to the corresponding ones from classical random walk. This might be surprising at first sight since quantum walks are phase coherent phenomenon. To explain the results in Figs. \ref{fig:ffrel}-\ref{fig:eeprob}, let us consider the state of the walker at $i$-th vertex to be $\psi_i(t) = a_i(t) ~ e^{i \theta_i(t)}$, with $a_i(t)$ and $\theta_i(t)$ being its instantaneous amplitude and phase. The cross-correlation between variable $X$ at vertices $i$ and $j$ can be defined as $C_{ij}(\tau) = \langle X_i(t) ~ X_j(t+\tau) \rangle$, where $\langle . \rangle$ represent time average, and $X_i=\theta_i$ or $X_i=z_i$. Also,  $C_{ii}(\tau) = \langle X_i(t) ~ X_i(t+\tau) \rangle$ represents auto-correlation at vertex $i$.

\begin{figure}
\centering
\includegraphics[width=1.0\linewidth]{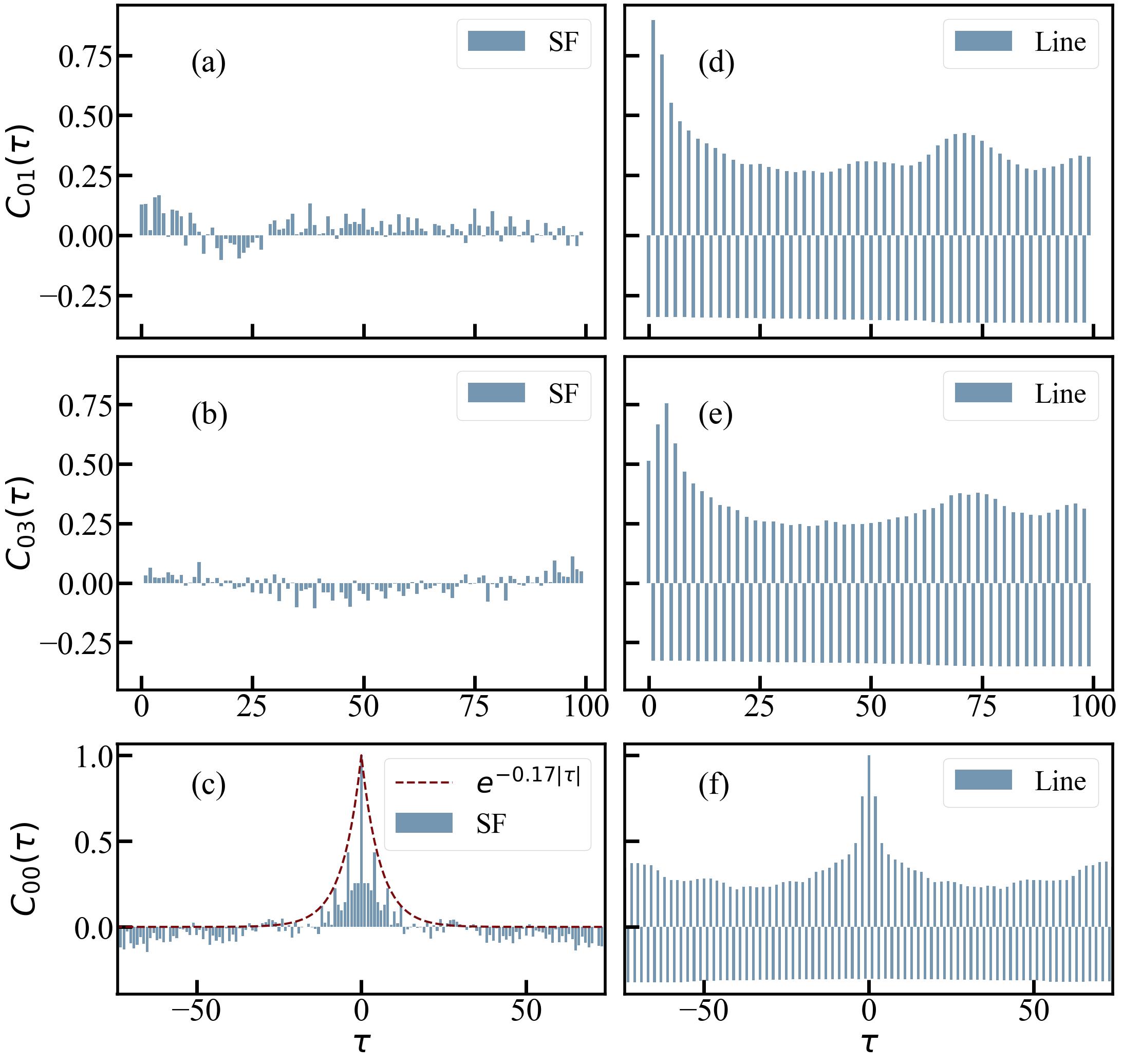}
\caption{Correlation $C_{ij}(\tau)$ between time-series of squared probability amplitudes at node $i$ and $j$; (a)-(c) for DTQW on a SF graph of 100 nodes, and (d)-(e) for DTQW on a periodic 1D lattice of size 100. The plots (a) and (d) show that for two adjacent nodes, while (b) and (e) show the correlations between two nodes separated by 2 edges. (c) and (f) show auto-correlation at a node labeled $0$ in respective graphs.}
\label{fig:corrz}
\end{figure}

Let us first examine the correlations in $z$ on SF graph, and denote the shortest path between vertices $i$ and $j$ as $d_{ij}$. Figure \ref{fig:corrz}(a) shows cross-correlation $C_{01}(\tau)$ between neighboring vertices labeled 0 ($k_0=22$) and 1 ($k_1=17$) with $d_{01}=1$, while Fig. \ref{fig:corrz}(b) shows $C_{03}(\tau)$ between vertices 0 and 3 ($k_3=21$) separated by $d_{03}=2$. Figure \ref{fig:corrz}(a,b) does not reveal any significant correlation, and $C_{ij}(\tau) \to 0$ as the shortest path between $i$ and $j$ gets longer. The auto-correlation $C_{00}(\tau)$ in Fig. \ref{fig:corrz}(c) shows exponential decay of correlation profile implying that $z(t)$ behaves as though it is a stochastic process (see Fig. \ref{fig:spread}(d)) despite deterministic quantum evolution. In the disordered SF lattice, presence of a variable number of neighbors therefore enhances the de-cohering effect, and effectively, induces randomness in $z(t)$.
\begin{figure}
\centering
\includegraphics[width=1\linewidth]{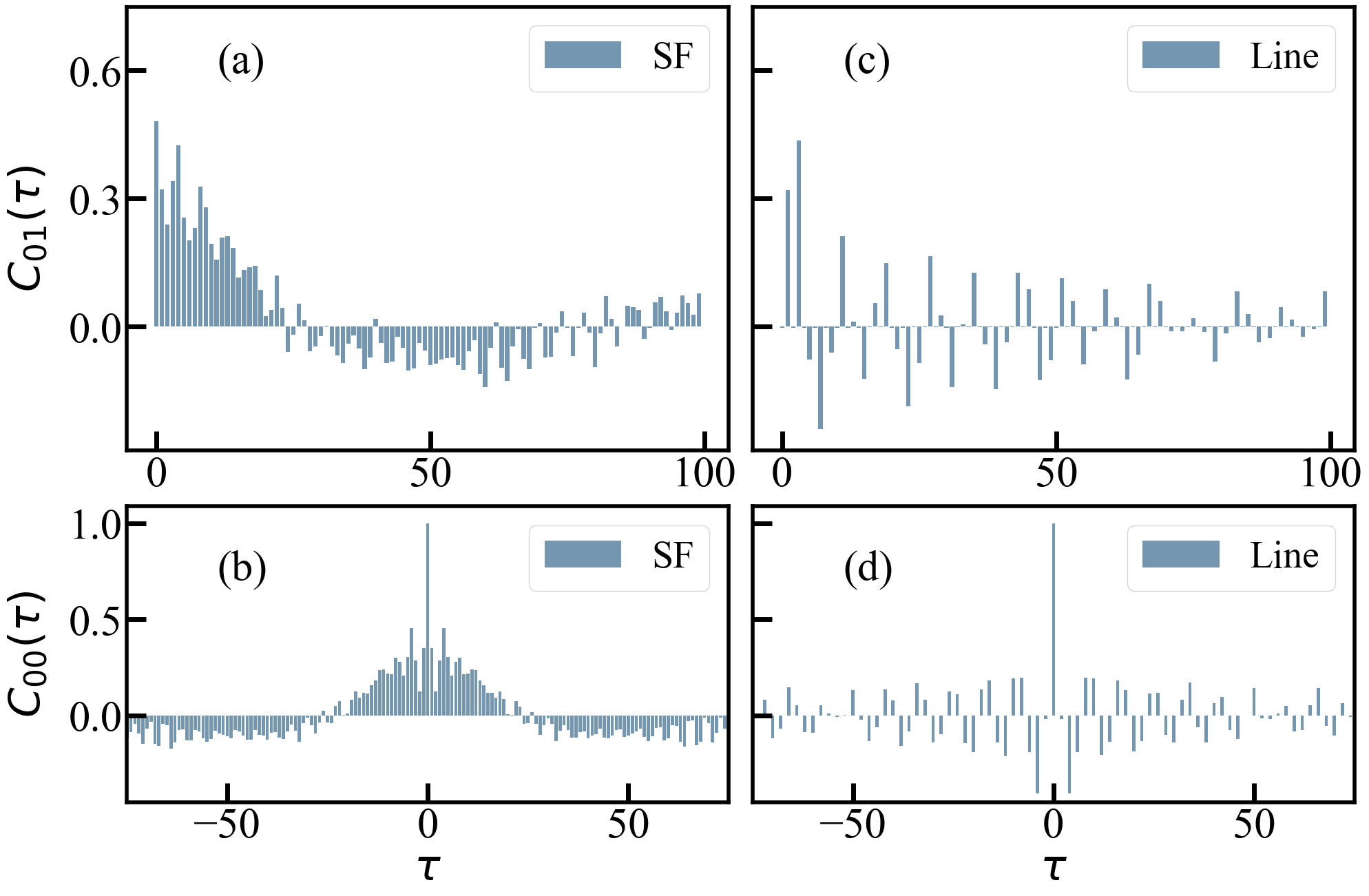}
\caption{Correlation $C_{ij}(\tau)$ between time-series of phases of the probability amplitudes at node $i$ and $j$ for DTQW on a SF graph of 100 nodes (a) and (b), and for DTQW on a periodic 1D lattice of size 100 (c)-(d). The two nodes $0$ and $1$ are adjacent. (b) and (d) show auto-correlation of the phases.}
\label{fig:corr_phase}
\end{figure}
Now, we examine the correlation in $z$ on the ring lattice with $k_i=2$ for all vertices. Figure \ref{fig:corrz}(d,e) reveals significant correlations between two pairs of vertices $i=0, j=1$ (with $d_{01}=1$ and $i=0,j=3$ (with $d_{03}=2$) and they are sustained for a long time. The periodicity and homogeneity of the lattice suppresses random scattering of the evolving wave-packet and hence large correlations are realized. Figure \ref{fig:corrz}(f) shows that $C_{00}(\tau)$ saturates at some non-zero value, consistent with what would be anticipated based on Fig. \ref{fig:spread}(d). Note that the extreme event probability for QWs, shown in Fig. \ref{fig:eeprob}, depends on dynamics of $z_i(t)$ on $i$-th vertex, irrespective of the nature of dynamics on other lattice points. This property is identical to that of corresponding classical walks on lattices. Based on Fig. \ref{fig:corrz}, we might infer that the local dynamics of $z(t)$, on SF graph, behaves similar to a nearly-uncorrelated stochastic process with an exponential auto-correlation function, while on a ring graph $z(t)$ behaves like a correlated stochastic process. Hence, the extreme event probability in Fig. \ref{fig:eeprob} is qualitatively similar to the corresponding classical walk.

Now, we examine the dynamics of phases $\theta$. Since QW evolution maintains some level of phase coherence with neighboring vertices, we expect that the phases on neighboring vertices might be correlated. This is borne out in Fig. \ref{fig:corr_phase}(a,c), for QW on SF and ring graphs, which reveal strong phase correlations among neighboring vertices. However, the autocorrelation shown in Fig. \ref{fig:corr_phase}(b,d) decays exponentially (on SF graph) and is faster than exponential decay (on ring graph). These observations can be summarized as follows : on a SF and ring lattice, the phases $\theta(t)$ between a pair of neighboring vertices are correlated, but $C_{ii}$ at $i$-th vertex has a stochastic character with fast decay of correlations. Even though QW is a phase coherent phenomenon and maintains phase coherence across nodes, but when focused on any particular node, phase coherence is not highly pronounced and the dynamics of $z$ is stochastic. This implies that, even for extremes in phase variable $\theta$, we could have obtained similar results as shown in Fig. \ref{fig:eeprob} for $z$.

{\sl Summary} : Extreme events arising from multiple classical random walkers on complex networks had been studied earlier. In this work, we set up the framework for defining extreme events of quantum walks on disordered lattices, and analyze the occurrence probability for extreme events on vertices of scale-free graphs and ring lattices. Though we expect quantum walks to be a coherent phenomenon and must deviate from classical random walks, surprisingly, at least as far as extreme events are concerned the results have qualitative similarity to that of classical random walks. In particular, contrary to expectations, the extreme event probability on a scale-free graph is larger for small degree vertices than for the hubs on the graph. Further, extreme event probability for quantum walks scale as a function of threshold used to identify extreme events. This work shows that extremes can arise even in phase coherent phenomenon with linear quantum dynamics.

{\sl Acknowledgment.}--- Nisarg Vyas would like to thank TCS Foundation for financial support with the TCS RSP Fellowship (2024-2028). 
\nocite{aharonov}
\nocite{KollarGilyKissStef2020}
\nocite{InuiKonishiKonno2004}

\bibliography{qwee}

\providecommand{\noopsort}[1]{}\providecommand{\singleletter}[1]{#1}%
\begin{thebibliography}{51}%
\makeatletter
\providecommand \@ifxundefined [1]{%
 \@ifx{#1\undefined}
}%
\providecommand \@ifnum [1]{%
 \ifnum #1\expandafter \@firstoftwo
 \else \expandafter \@secondoftwo
 \fi
}%
\providecommand \@ifx [1]{%
 \ifx #1\expandafter \@firstoftwo
 \else \expandafter \@secondoftwo
 \fi
}%
\providecommand \natexlab [1]{#1}%
\providecommand \enquote  [1]{``#1''}%
\providecommand \bibnamefont  [1]{#1}%
\providecommand \bibfnamefont [1]{#1}%
\providecommand \citenamefont [1]{#1}%
\providecommand \href@noop [0]{\@secondoftwo}%
\providecommand \href [0]{\begingroup \@sanitize@url \@href}%
\providecommand \@href[1]{\@@startlink{#1}\@@href}%
\providecommand \@@href[1]{\endgroup#1\@@endlink}%
\providecommand \@sanitize@url [0]{\catcode `\\12\catcode `\$12\catcode
  `\&12\catcode `\#12\catcode `\^12\catcode `\_12\catcode `\%12\relax}%
\providecommand \@@startlink[1]{}%
\providecommand \@@endlink[0]{}%
\providecommand \url  [0]{\begingroup\@sanitize@url \@url }%
\providecommand \@url [1]{\endgroup\@href {#1}{\urlprefix }}%
\providecommand \urlprefix  [0]{URL }%
\providecommand \Eprint [0]{\href }%
\providecommand \doibase [0]{https://doi.org/}%
\providecommand \selectlanguage [0]{\@gobble}%
\providecommand \bibinfo  [0]{\@secondoftwo}%
\providecommand \bibfield  [0]{\@secondoftwo}%
\providecommand \translation [1]{[#1]}%
\providecommand \BibitemOpen [0]{}%
\providecommand \bibitemStop [0]{}%
\providecommand \bibitemNoStop [0]{.\EOS\space}%
\providecommand \EOS [0]{\spacefactor3000\relax}%
\providecommand \BibitemShut  [1]{\csname bibitem#1\endcsname}%
\let\auto@bib@innerbib\@empty
\bibitem [{\citenamefont {Aharonov}\ \emph {et~al.}(1993)\citenamefont
  {Aharonov}, \citenamefont {Davidovich},\ and\ \citenamefont
  {Zagury}}]{qrw_aharonov}%
  \BibitemOpen
  \bibfield  {author} {\bibinfo {author} {\bibfnamefont {Y.}~\bibnamefont
  {Aharonov}}, \bibinfo {author} {\bibfnamefont {L.}~\bibnamefont
  {Davidovich}},\ and\ \bibinfo {author} {\bibfnamefont {N.}~\bibnamefont
  {Zagury}},\ }\bibfield  {title} {\bibinfo {title} {Quantum random walks},\
  }\href {https://doi.org/10.1103/PhysRevA.48.1687} {\bibfield  {journal}
  {\bibinfo  {journal} {Phys. Rev. A}\ }\textbf {\bibinfo {volume} {48}},\
  \bibinfo {pages} {1687} (\bibinfo {year} {1993})}\BibitemShut {NoStop}%
\bibitem [{\citenamefont {Ambainis}(2003)}]{Ambainis_quant_algo}%
  \BibitemOpen
  \bibfield  {author} {\bibinfo {author} {\bibfnamefont {A.}~\bibnamefont
  {Ambainis}},\ }\bibfield  {title} {\bibinfo {title} {Quantum walks and their
  algorithmic applications},\ }\href
  {https://doi.org/10.1142/S0219749903000383} {\bibfield  {journal} {\bibinfo
  {journal} {International Journal of Quantum Information}\ }\textbf {\bibinfo
  {volume} {01}},\ \bibinfo {pages} {507} (\bibinfo {year} {2003})},\ \Eprint
  {https://arxiv.org/abs/https://doi.org/10.1142/S0219749903000383}
  {https://doi.org/10.1142/S0219749903000383} \BibitemShut {NoStop}%
\bibitem [{\citenamefont {Childs}\ \emph {et~al.}(2003)\citenamefont {Childs},
  \citenamefont {Cleve}, \citenamefont {Deotto}, \citenamefont {Farhi},
  \citenamefont {Gutmann},\ and\ \citenamefont
  {Spielman}}]{childs2003_expoHit}%
  \BibitemOpen
  \bibfield  {author} {\bibinfo {author} {\bibfnamefont {A.~M.}\ \bibnamefont
  {Childs}}, \bibinfo {author} {\bibfnamefont {R.}~\bibnamefont {Cleve}},
  \bibinfo {author} {\bibfnamefont {E.}~\bibnamefont {Deotto}}, \bibinfo
  {author} {\bibfnamefont {E.}~\bibnamefont {Farhi}}, \bibinfo {author}
  {\bibfnamefont {S.}~\bibnamefont {Gutmann}},\ and\ \bibinfo {author}
  {\bibfnamefont {D.~A.}\ \bibnamefont {Spielman}},\ }\bibfield  {title}
  {\bibinfo {title} {Exponential algorithmic speedup by a quantum walk},\ }in\
  \href {https://doi.org/10.1145/780542.780552} {\emph {\bibinfo {booktitle}
  {Proceedings of the Thirty-Fifth Annual ACM Symposium on Theory of
  Computing}}},\ \bibinfo {series and number} {STOC '03}\ (\bibinfo
  {publisher} {Association for Computing Machinery},\ \bibinfo {address} {New
  York, NY, USA},\ \bibinfo {year} {2003})\ p.\ \bibinfo {pages}
  {59–68}\BibitemShut {NoStop}%
\bibitem [{\citenamefont {Kempe}(2002)}]{Kempe2002_expoHit}%
  \BibitemOpen
  \bibfield  {author} {\bibinfo {author} {\bibfnamefont {J.}~\bibnamefont
  {Kempe}},\ }\href {https://arxiv.org/abs/quant-ph/0205083} {\bibinfo {title}
  {Quantum random walks hit exponentially faster}} (\bibinfo {year} {2002}),\
  \Eprint {https://arxiv.org/abs/quant-ph/0205083} {arXiv:quant-ph/0205083
  [quant-ph]} \BibitemShut {NoStop}%
\bibitem [{\citenamefont {Farhi}\ and\ \citenamefont
  {Gutmann}(1997)}]{Farhi1997_expoHit}%
  \BibitemOpen
  \bibfield  {author} {\bibinfo {author} {\bibfnamefont {E.}~\bibnamefont
  {Farhi}}\ and\ \bibinfo {author} {\bibfnamefont {S.}~\bibnamefont
  {Gutmann}},\ }\bibfield  {title} {\bibinfo {title} {Quantum computation and
  decision trees},\ }\href {https://api.semanticscholar.org/CorpusID:1439479}
  {\bibfield  {journal} {\bibinfo  {journal} {Physical Review A}\ }\textbf
  {\bibinfo {volume} {58}},\ \bibinfo {pages} {915} (\bibinfo {year}
  {1997})}\BibitemShut {NoStop}%
\bibitem [{\citenamefont {Mülken}\ and\ \citenamefont
  {Blumen}(2006)}]{M_lken_2006}%
  \BibitemOpen
  \bibfield  {author} {\bibinfo {author} {\bibfnamefont {O.}~\bibnamefont
  {Mülken}}\ and\ \bibinfo {author} {\bibfnamefont {A.}~\bibnamefont
  {Blumen}},\ }\bibfield  {title} {\bibinfo {title} {Efficiency of quantum and
  classical transport on graphs},\ }\bibfield  {journal} {\bibinfo  {journal}
  {Physical Review E}\ }\textbf {\bibinfo {volume} {73}},\ \href
  {https://doi.org/10.1103/physreve.73.066117} {10.1103/physreve.73.066117}
  (\bibinfo {year} {2006})\BibitemShut {NoStop}%
\bibitem [{\citenamefont {Ambainis}(2007)}]{ambainis2014_distinctness}%
  \BibitemOpen
  \bibfield  {author} {\bibinfo {author} {\bibfnamefont {A.}~\bibnamefont
  {Ambainis}},\ }\bibfield  {title} {\bibinfo {title} {Quantum walk algorithm
  for element distinctness},\ }\href
  {https://doi.org/10.1137/S0097539705447311} {\bibfield  {journal} {\bibinfo
  {journal} {SIAM Journal on Computing}\ }\textbf {\bibinfo {volume} {37}},\
  \bibinfo {pages} {210} (\bibinfo {year} {2007})},\ \Eprint
  {https://arxiv.org/abs/https://doi.org/10.1137/S0097539705447311}
  {https://doi.org/10.1137/S0097539705447311} \BibitemShut {NoStop}%
\bibitem [{\citenamefont {Magniez}\ \emph {et~al.}(2007)\citenamefont
  {Magniez}, \citenamefont {Santha},\ and\ \citenamefont
  {Szegedy}}]{magniez2005_triangleFind}%
  \BibitemOpen
  \bibfield  {author} {\bibinfo {author} {\bibfnamefont {F.}~\bibnamefont
  {Magniez}}, \bibinfo {author} {\bibfnamefont {M.}~\bibnamefont {Santha}},\
  and\ \bibinfo {author} {\bibfnamefont {M.}~\bibnamefont {Szegedy}},\
  }\bibfield  {title} {\bibinfo {title} {Quantum algorithms for the triangle
  problem},\ }\href {https://doi.org/10.1137/050643684} {\bibfield  {journal}
  {\bibinfo  {journal} {SIAM Journal on Computing}\ }\textbf {\bibinfo {volume}
  {37}},\ \bibinfo {pages} {413} (\bibinfo {year} {2007})},\ \Eprint
  {https://arxiv.org/abs/https://doi.org/10.1137/050643684}
  {https://doi.org/10.1137/050643684} \BibitemShut {NoStop}%
\bibitem [{\citenamefont {Yan}\ \emph {et~al.}(2019)\citenamefont {Yan},
  \citenamefont {Zhang}, \citenamefont {Gong}, \citenamefont {Wu},
  \citenamefont {Zheng}, \citenamefont {Li}, \citenamefont {Wang},
  \citenamefont {Liang}, \citenamefont {Lin}, \citenamefont {Xu}, \citenamefont
  {Guo}, \citenamefont {Sun}, \citenamefont {Peng}, \citenamefont {Xia},
  \citenamefont {Deng}, \citenamefont {Rong}, \citenamefont {You},
  \citenamefont {Nori}, \citenamefont {Fan}, \citenamefont {Zhu},\ and\
  \citenamefont {Pan}}]{YanZhaGon2019}%
  \BibitemOpen
  \bibfield  {author} {\bibinfo {author} {\bibfnamefont {Z.}~\bibnamefont
  {Yan}}, \bibinfo {author} {\bibfnamefont {Y.-R.}\ \bibnamefont {Zhang}},
  \bibinfo {author} {\bibfnamefont {M.}~\bibnamefont {Gong}}, \bibinfo {author}
  {\bibfnamefont {Y.}~\bibnamefont {Wu}}, \bibinfo {author} {\bibfnamefont
  {Y.}~\bibnamefont {Zheng}}, \bibinfo {author} {\bibfnamefont
  {S.}~\bibnamefont {Li}}, \bibinfo {author} {\bibfnamefont {C.}~\bibnamefont
  {Wang}}, \bibinfo {author} {\bibfnamefont {F.}~\bibnamefont {Liang}},
  \bibinfo {author} {\bibfnamefont {J.}~\bibnamefont {Lin}}, \bibinfo {author}
  {\bibfnamefont {Y.}~\bibnamefont {Xu}}, \bibinfo {author} {\bibfnamefont
  {C.}~\bibnamefont {Guo}}, \bibinfo {author} {\bibfnamefont {L.}~\bibnamefont
  {Sun}}, \bibinfo {author} {\bibfnamefont {C.-Z.}\ \bibnamefont {Peng}},
  \bibinfo {author} {\bibfnamefont {K.}~\bibnamefont {Xia}}, \bibinfo {author}
  {\bibfnamefont {H.}~\bibnamefont {Deng}}, \bibinfo {author} {\bibfnamefont
  {H.}~\bibnamefont {Rong}}, \bibinfo {author} {\bibfnamefont {J.~Q.}\
  \bibnamefont {You}}, \bibinfo {author} {\bibfnamefont {F.}~\bibnamefont
  {Nori}}, \bibinfo {author} {\bibfnamefont {H.}~\bibnamefont {Fan}}, \bibinfo
  {author} {\bibfnamefont {X.}~\bibnamefont {Zhu}},\ and\ \bibinfo {author}
  {\bibfnamefont {J.-W.}\ \bibnamefont {Pan}},\ }\bibfield  {title} {\bibinfo
  {title} {Strongly correlated quantum walks with a 12-qubit superconducting
  processor},\ }\href {https://doi.org/10.1126/science.aaw1611} {\bibfield
  {journal} {\bibinfo  {journal} {Science}\ }\textbf {\bibinfo {volume}
  {364}},\ \bibinfo {pages} {753} (\bibinfo {year} {2019})}\BibitemShut
  {NoStop}%
\bibitem [{\citenamefont {Flamini}\ \emph {et~al.}(2018)\citenamefont
  {Flamini}, \citenamefont {Spagnolo},\ and\ \citenamefont
  {Sciarrino}}]{FlaSpaSci2018}%
  \BibitemOpen
  \bibfield  {author} {\bibinfo {author} {\bibfnamefont {F.}~\bibnamefont
  {Flamini}}, \bibinfo {author} {\bibfnamefont {N.}~\bibnamefont {Spagnolo}},\
  and\ \bibinfo {author} {\bibfnamefont {F.}~\bibnamefont {Sciarrino}},\
  }\bibfield  {title} {\bibinfo {title} {Photonic quantum information
  processing: a review},\ }\href {https://doi.org/10.1088/1361-6633/aad5b2}
  {\bibfield  {journal} {\bibinfo  {journal} {Reports on Progress in Physics}\
  }\textbf {\bibinfo {volume} {82}},\ \bibinfo {pages} {016001} (\bibinfo
  {year} {2018})}\BibitemShut {NoStop}%
\bibitem [{\citenamefont {Broome}\ \emph {et~al.}(2010)\citenamefont {Broome},
  \citenamefont {Fedrizzi}, \citenamefont {Lanyon}, \citenamefont {Kassal},
  \citenamefont {Aspuru-Guzik},\ and\ \citenamefont {White}}]{BroFedLan2010}%
  \BibitemOpen
  \bibfield  {author} {\bibinfo {author} {\bibfnamefont {M.~A.}\ \bibnamefont
  {Broome}}, \bibinfo {author} {\bibfnamefont {A.}~\bibnamefont {Fedrizzi}},
  \bibinfo {author} {\bibfnamefont {B.~P.}\ \bibnamefont {Lanyon}}, \bibinfo
  {author} {\bibfnamefont {I.}~\bibnamefont {Kassal}}, \bibinfo {author}
  {\bibfnamefont {A.}~\bibnamefont {Aspuru-Guzik}},\ and\ \bibinfo {author}
  {\bibfnamefont {A.~G.}\ \bibnamefont {White}},\ }\bibfield  {title} {\bibinfo
  {title} {Discrete single-photon quantum walks with tunable decoherence},\
  }\href {https://doi.org/10.1103/PhysRevLett.104.153602} {\bibfield  {journal}
  {\bibinfo  {journal} {Phys. Rev. Lett.}\ }\textbf {\bibinfo {volume} {104}},\
  \bibinfo {pages} {153602} (\bibinfo {year} {2010})}\BibitemShut {NoStop}%
\bibitem [{\citenamefont {Schreiber}\ \emph {et~al.}(2011)\citenamefont
  {Schreiber}, \citenamefont {Cassemiro}, \citenamefont
  {Poto\ifmmode~\check{c}\else \v{c}\fi{}ek}, \citenamefont {G\'abris},
  \citenamefont {Jex},\ and\ \citenamefont {Silberhorn}}]{SchCasPot2011}%
  \BibitemOpen
  \bibfield  {author} {\bibinfo {author} {\bibfnamefont {A.}~\bibnamefont
  {Schreiber}}, \bibinfo {author} {\bibfnamefont {K.~N.}\ \bibnamefont
  {Cassemiro}}, \bibinfo {author} {\bibfnamefont {V.}~\bibnamefont
  {Poto\ifmmode~\check{c}\else \v{c}\fi{}ek}}, \bibinfo {author} {\bibfnamefont
  {A.}~\bibnamefont {G\'abris}}, \bibinfo {author} {\bibfnamefont
  {I.}~\bibnamefont {Jex}},\ and\ \bibinfo {author} {\bibfnamefont
  {C.}~\bibnamefont {Silberhorn}},\ }\bibfield  {title} {\bibinfo {title}
  {Decoherence and disorder in quantum walks: From ballistic spread to
  localization},\ }\href {https://doi.org/10.1103/PhysRevLett.106.180403}
  {\bibfield  {journal} {\bibinfo  {journal} {Phys. Rev. Lett.}\ }\textbf
  {\bibinfo {volume} {106}},\ \bibinfo {pages} {180403} (\bibinfo {year}
  {2011})}\BibitemShut {NoStop}%
\bibitem [{\citenamefont {Sansoni}\ \emph {et~al.}(2012)\citenamefont
  {Sansoni}, \citenamefont {Sciarrino}, \citenamefont {Vallone}, \citenamefont
  {Mataloni}, \citenamefont {Crespi}, \citenamefont {Ramponi},\ and\
  \citenamefont {Osellame}}]{SanSciVal2012}%
  \BibitemOpen
  \bibfield  {author} {\bibinfo {author} {\bibfnamefont {L.}~\bibnamefont
  {Sansoni}}, \bibinfo {author} {\bibfnamefont {F.}~\bibnamefont {Sciarrino}},
  \bibinfo {author} {\bibfnamefont {G.}~\bibnamefont {Vallone}}, \bibinfo
  {author} {\bibfnamefont {P.}~\bibnamefont {Mataloni}}, \bibinfo {author}
  {\bibfnamefont {A.}~\bibnamefont {Crespi}}, \bibinfo {author} {\bibfnamefont
  {R.}~\bibnamefont {Ramponi}},\ and\ \bibinfo {author} {\bibfnamefont
  {R.}~\bibnamefont {Osellame}},\ }\bibfield  {title} {\bibinfo {title}
  {Two-particle bosonic-fermionic quantum walk via integrated photonics},\
  }\href {https://doi.org/10.1103/PhysRevLett.108.010502} {\bibfield  {journal}
  {\bibinfo  {journal} {Phys. Rev. Lett.}\ }\textbf {\bibinfo {volume} {108}},\
  \bibinfo {pages} {010502} (\bibinfo {year} {2012})}\BibitemShut {NoStop}%
\bibitem [{\citenamefont {Crespi}\ \emph {et~al.}(2013)\citenamefont {Crespi},
  \citenamefont {Osellame}, \citenamefont {Ramponi}, \citenamefont
  {Giovannetti}, \citenamefont {Fazio}, \citenamefont {Sansoni}, \citenamefont
  {De~Nicola}, \citenamefont {Sciarrino},\ and\ \citenamefont
  {Mataloni}}]{CreOseRam2013}%
  \BibitemOpen
  \bibfield  {author} {\bibinfo {author} {\bibfnamefont {A.}~\bibnamefont
  {Crespi}}, \bibinfo {author} {\bibfnamefont {R.}~\bibnamefont {Osellame}},
  \bibinfo {author} {\bibfnamefont {R.}~\bibnamefont {Ramponi}}, \bibinfo
  {author} {\bibfnamefont {V.}~\bibnamefont {Giovannetti}}, \bibinfo {author}
  {\bibfnamefont {R.}~\bibnamefont {Fazio}}, \bibinfo {author} {\bibfnamefont
  {L.}~\bibnamefont {Sansoni}}, \bibinfo {author} {\bibfnamefont
  {F.}~\bibnamefont {De~Nicola}}, \bibinfo {author} {\bibfnamefont
  {F.}~\bibnamefont {Sciarrino}},\ and\ \bibinfo {author} {\bibfnamefont
  {P.}~\bibnamefont {Mataloni}},\ }\bibfield  {title} {\bibinfo {title}
  {{Anderson localization of entangled photons in an integrated quantum
  walk}},\ }\href {https://doi.org/10.1038/nphoton.2013.26} {\bibfield
  {journal} {\bibinfo  {journal} {Nat. Photonics}\ }\textbf {\bibinfo {volume}
  {7}},\ \bibinfo {pages} {322} (\bibinfo {year} {2013})}\BibitemShut {NoStop}%
\bibitem [{\citenamefont {Karski}\ \emph {et~al.}(2009)\citenamefont {Karski},
  \citenamefont {Förster}, \citenamefont {Choi}, \citenamefont {Steffen},
  \citenamefont {Alt}, \citenamefont {Meschede},\ and\ \citenamefont
  {Widera}}]{MicLeoJai2009}%
  \BibitemOpen
  \bibfield  {author} {\bibinfo {author} {\bibfnamefont {M.}~\bibnamefont
  {Karski}}, \bibinfo {author} {\bibfnamefont {L.}~\bibnamefont {Förster}},
  \bibinfo {author} {\bibfnamefont {J.-M.}\ \bibnamefont {Choi}}, \bibinfo
  {author} {\bibfnamefont {A.}~\bibnamefont {Steffen}}, \bibinfo {author}
  {\bibfnamefont {W.}~\bibnamefont {Alt}}, \bibinfo {author} {\bibfnamefont
  {D.}~\bibnamefont {Meschede}},\ and\ \bibinfo {author} {\bibfnamefont
  {A.}~\bibnamefont {Widera}},\ }\bibfield  {title} {\bibinfo {title} {Quantum
  walk in position space with single optically trapped atoms},\ }\href
  {https://doi.org/10.1126/science.1174436} {\bibfield  {journal} {\bibinfo
  {journal} {Science}\ }\textbf {\bibinfo {volume} {325}},\ \bibinfo {pages}
  {174} (\bibinfo {year} {2009})}\BibitemShut {NoStop}%
\bibitem [{\citenamefont {Z\"ahringer}\ \emph {et~al.}(2010)\citenamefont
  {Z\"ahringer}, \citenamefont {Kirchmair}, \citenamefont {Gerritsma},
  \citenamefont {Solano}, \citenamefont {Blatt},\ and\ \citenamefont
  {Roos}}]{ZahKirGer2010}%
  \BibitemOpen
  \bibfield  {author} {\bibinfo {author} {\bibfnamefont {F.}~\bibnamefont
  {Z\"ahringer}}, \bibinfo {author} {\bibfnamefont {G.}~\bibnamefont
  {Kirchmair}}, \bibinfo {author} {\bibfnamefont {R.}~\bibnamefont
  {Gerritsma}}, \bibinfo {author} {\bibfnamefont {E.}~\bibnamefont {Solano}},
  \bibinfo {author} {\bibfnamefont {R.}~\bibnamefont {Blatt}},\ and\ \bibinfo
  {author} {\bibfnamefont {C.~F.}\ \bibnamefont {Roos}},\ }\bibfield  {title}
  {\bibinfo {title} {Realization of a quantum walk with one and two trapped
  ions},\ }\href {https://doi.org/10.1103/PhysRevLett.104.100503} {\bibfield
  {journal} {\bibinfo  {journal} {Phys. Rev. Lett.}\ }\textbf {\bibinfo
  {volume} {104}},\ \bibinfo {pages} {100503} (\bibinfo {year}
  {2010})}\BibitemShut {NoStop}%
\bibitem [{\citenamefont {Preiss}\ \emph {et~al.}(2015)\citenamefont {Preiss},
  \citenamefont {Ma}, \citenamefont {Tai}, \citenamefont {Lukin}, \citenamefont
  {Rispoli}, \citenamefont {Zupancic}, \citenamefont {Lahini}, \citenamefont
  {Islam},\ and\ \citenamefont {Greiner}}]{PreMaTai2015}%
  \BibitemOpen
  \bibfield  {author} {\bibinfo {author} {\bibfnamefont {P.~M.}\ \bibnamefont
  {Preiss}}, \bibinfo {author} {\bibfnamefont {R.}~\bibnamefont {Ma}}, \bibinfo
  {author} {\bibfnamefont {M.~E.}\ \bibnamefont {Tai}}, \bibinfo {author}
  {\bibfnamefont {A.}~\bibnamefont {Lukin}}, \bibinfo {author} {\bibfnamefont
  {M.}~\bibnamefont {Rispoli}}, \bibinfo {author} {\bibfnamefont
  {P.}~\bibnamefont {Zupancic}}, \bibinfo {author} {\bibfnamefont
  {Y.}~\bibnamefont {Lahini}}, \bibinfo {author} {\bibfnamefont
  {R.}~\bibnamefont {Islam}},\ and\ \bibinfo {author} {\bibfnamefont
  {M.}~\bibnamefont {Greiner}},\ }\bibfield  {title} {\bibinfo {title}
  {Strongly correlated quantum walks in optical lattices},\ }\href
  {https://doi.org/10.1126/science.1260364} {\bibfield  {journal} {\bibinfo
  {journal} {Science}\ }\textbf {\bibinfo {volume} {347}},\ \bibinfo {pages}
  {1229} (\bibinfo {year} {2015})}\BibitemShut {NoStop}%
\bibitem [{\citenamefont {Duda}\ \emph {et~al.}(2023)\citenamefont {Duda},
  \citenamefont {Ivaki}, \citenamefont {Sahlberg}, \citenamefont
  {P\"oyh\"onen},\ and\ \citenamefont {Ojanen}}]{DudIvaSah2023}%
  \BibitemOpen
  \bibfield  {author} {\bibinfo {author} {\bibfnamefont {R.}~\bibnamefont
  {Duda}}, \bibinfo {author} {\bibfnamefont {M.~N.}\ \bibnamefont {Ivaki}},
  \bibinfo {author} {\bibfnamefont {I.}~\bibnamefont {Sahlberg}}, \bibinfo
  {author} {\bibfnamefont {K.}~\bibnamefont {P\"oyh\"onen}},\ and\ \bibinfo
  {author} {\bibfnamefont {T.}~\bibnamefont {Ojanen}},\ }\bibfield  {title}
  {\bibinfo {title} {Quantum walks on random lattices: Diffusion, localization,
  and the absence of parametric quantum speedup},\ }\href
  {https://doi.org/10.1103/PhysRevResearch.5.023150} {\bibfield  {journal}
  {\bibinfo  {journal} {Phys. Rev. Res.}\ }\textbf {\bibinfo {volume} {5}},\
  \bibinfo {pages} {023150} (\bibinfo {year} {2023})}\BibitemShut {NoStop}%
\bibitem [{\citenamefont {Sharma}\ and\ \citenamefont
  {Boettcher}(2022)}]{ShaBoe2022}%
  \BibitemOpen
  \bibfield  {author} {\bibinfo {author} {\bibfnamefont {R.}~\bibnamefont
  {Sharma}}\ and\ \bibinfo {author} {\bibfnamefont {S.}~\bibnamefont
  {Boettcher}},\ }\bibfield  {title} {\bibinfo {title} {Transport and
  localization in quantum walks on a random hierarchy of barriers},\ }\href
  {https://doi.org/10.1088/1751-8121/ac7117} {\bibfield  {journal} {\bibinfo
  {journal} {Journal of Physics A: Mathematical and Theoretical}\ }\textbf
  {\bibinfo {volume} {55}},\ \bibinfo {pages} {264001} (\bibinfo {year}
  {2022})}\BibitemShut {NoStop}%
\bibitem [{\citenamefont {Li}\ \emph {et~al.}(2013)\citenamefont {Li},
  \citenamefont {Izaac},\ and\ \citenamefont {Wang}}]{LiIzaWan2013}%
  \BibitemOpen
  \bibfield  {author} {\bibinfo {author} {\bibfnamefont {Z.~J.}\ \bibnamefont
  {Li}}, \bibinfo {author} {\bibfnamefont {J.~A.}\ \bibnamefont {Izaac}},\ and\
  \bibinfo {author} {\bibfnamefont {J.~B.}\ \bibnamefont {Wang}},\ }\bibfield
  {title} {\bibinfo {title} {Position-defect-induced reflection, trapping,
  transmission, and resonance in quantum walks},\ }\href
  {https://doi.org/10.1103/PhysRevA.87.012314} {\bibfield  {journal} {\bibinfo
  {journal} {Phys. Rev. A}\ }\textbf {\bibinfo {volume} {87}},\ \bibinfo
  {pages} {012314} (\bibinfo {year} {2013})}\BibitemShut {NoStop}%
\bibitem [{\citenamefont {Yao}\ and\ \citenamefont {Wald}(2023)}]{YaoWald2023}%
  \BibitemOpen
  \bibfield  {author} {\bibinfo {author} {\bibfnamefont {L.~H.}\ \bibnamefont
  {Yao}}\ and\ \bibinfo {author} {\bibfnamefont {S.}~\bibnamefont {Wald}},\
  }\bibfield  {title} {\bibinfo {title} {Coined quantum walks on the line:
  Disorder, entanglement, and localization},\ }\href
  {https://doi.org/10.1103/PhysRevE.108.024139} {\bibfield  {journal} {\bibinfo
   {journal} {Phys. Rev. E}\ }\textbf {\bibinfo {volume} {108}},\ \bibinfo
  {pages} {024139} (\bibinfo {year} {2023})}\BibitemShut {NoStop}%
\bibitem [{\citenamefont {Zeng}\ and\ \citenamefont
  {Yong}(2017)}]{ZengYong2017}%
  \BibitemOpen
  \bibfield  {author} {\bibinfo {author} {\bibfnamefont {M.}~\bibnamefont
  {Zeng}}\ and\ \bibinfo {author} {\bibfnamefont {E.~H.}\ \bibnamefont
  {Yong}},\ }\bibfield  {title} {\bibinfo {title} {Discrete-time quantum walk
  with phase disorder: Localization and entanglement entropy},\ }\href
  {https://doi.org/10.1038/s41598-017-12077-0} {\bibfield  {journal} {\bibinfo
  {journal} {Scientific Reports}\ }\textbf {\bibinfo {volume} {7}},\ \bibinfo
  {pages} {12024} (\bibinfo {year} {2017})}\BibitemShut {NoStop}%
\bibitem [{\citenamefont {Sethna}(2021)}]{Sethna_emergent}%
  \BibitemOpen
  \bibfield  {author} {\bibinfo {author} {\bibfnamefont {J.~P.}\ \bibnamefont
  {Sethna}},\ }\bibfield  {title} {\bibinfo {title} {23random walks and
  emergent properties},\ }in\ \href
  {https://doi.org/10.1093/oso/9780198865247.003.0002} {\emph {\bibinfo
  {booktitle} {Statistical Mechanics: Entropy, Order Parameters, and
  Complexity}}}\ (\bibinfo  {publisher} {Oxford University Press},\ \bibinfo
  {year} {2021})\BibitemShut {NoStop}%
\bibitem [{\citenamefont {Figueiredo}\ and\ \citenamefont
  {Garetto}(2017)}]{FiGaretto2017}%
  \BibitemOpen
  \bibfield  {author} {\bibinfo {author} {\bibfnamefont {D.~R.}\ \bibnamefont
  {Figueiredo}}\ and\ \bibinfo {author} {\bibfnamefont {M.}~\bibnamefont
  {Garetto}},\ }\bibfield  {title} {\bibinfo {title} {{ On the Emergence of
  Shortest Paths by Reinforced Random Walks }},\ }\href
  {https://doi.org/10.1109/TNSE.2016.2618063} {\bibfield  {journal} {\bibinfo
  {journal} {IEEE Transactions on Network Science and Engineering}\ }\textbf
  {\bibinfo {volume} {4}},\ \bibinfo {pages} {55} (\bibinfo {year}
  {2017})}\BibitemShut {NoStop}%
\bibitem [{\citenamefont {Stuhrmann}\ and\ \citenamefont
  {Coghi}(2024)}]{SturhmannCoghi2024}%
  \BibitemOpen
  \bibfield  {author} {\bibinfo {author} {\bibfnamefont {D.~C.}\ \bibnamefont
  {Stuhrmann}}\ and\ \bibinfo {author} {\bibfnamefont {F.}~\bibnamefont
  {Coghi}},\ }\bibfield  {title} {\bibinfo {title} {Understanding random-walk
  dynamical phase coexistence through waiting times},\ }\href
  {https://doi.org/10.1103/PhysRevResearch.6.013077} {\bibfield  {journal}
  {\bibinfo  {journal} {Phys. Rev. Res.}\ }\textbf {\bibinfo {volume} {6}},\
  \bibinfo {pages} {013077} (\bibinfo {year} {2024})}\BibitemShut {NoStop}%
\bibitem [{\citenamefont {Lecheval}\ \emph {et~al.}(2024)\citenamefont
  {Lecheval}, \citenamefont {Robinson},\ and\ \citenamefont
  {Mann}}]{LechRobMann2024}%
  \BibitemOpen
  \bibfield  {author} {\bibinfo {author} {\bibfnamefont {V.}~\bibnamefont
  {Lecheval}}, \bibinfo {author} {\bibfnamefont {E.~J.~H.}\ \bibnamefont
  {Robinson}},\ and\ \bibinfo {author} {\bibfnamefont {R.~P.}\ \bibnamefont
  {Mann}},\ }\bibfield  {title} {\bibinfo {title} {Random walks with spatial
  and temporal resets can explain individual and colony-level searching
  patterns in ants},\ }\href {https://doi.org/10.1098/rsif.2024.0149}
  {\bibfield  {journal} {\bibinfo  {journal} {Journal of The Royal Society
  Interface}\ }\textbf {\bibinfo {volume} {21}},\ \bibinfo {pages} {20240149}
  (\bibinfo {year} {2024})}\BibitemShut {NoStop}%
\bibitem [{\citenamefont {Barlovi{\'c}}\ \emph {et~al.}(2001)\citenamefont
  {Barlovi{\'c}}, \citenamefont {Schadschneider},\ and\ \citenamefont
  {Schreckenberg}}]{BarloSchadShreck2001}%
  \BibitemOpen
  \bibfield  {author} {\bibinfo {author} {\bibfnamefont {R.}~\bibnamefont
  {Barlovi{\'c}}}, \bibinfo {author} {\bibfnamefont {A.}~\bibnamefont
  {Schadschneider}},\ and\ \bibinfo {author} {\bibfnamefont {M.}~\bibnamefont
  {Schreckenberg}},\ }\bibfield  {title} {\bibinfo {title} {Random walk theory
  of jamming in a cellular automaton model for traffic flow},\ }\href
  {https://api.semanticscholar.org/CorpusID:15540877} {\bibfield  {journal}
  {\bibinfo  {journal} {Physica A-statistical Mechanics and Its Applications}\
  }\textbf {\bibinfo {volume} {294}},\ \bibinfo {pages} {525} (\bibinfo {year}
  {2001})}\BibitemShut {NoStop}%
\bibitem [{\citenamefont {Gandhi}\ and\ \citenamefont
  {Santhanam}(2022)}]{GandhiSanthanam2022}%
  \BibitemOpen
  \bibfield  {author} {\bibinfo {author} {\bibfnamefont {G.}~\bibnamefont
  {Gandhi}}\ and\ \bibinfo {author} {\bibfnamefont {M.~S.}\ \bibnamefont
  {Santhanam}},\ }\bibfield  {title} {\bibinfo {title} {Biased random walkers
  and extreme events on the edges of complex networks},\ }\href
  {https://doi.org/10.1103/PhysRevE.105.014315} {\bibfield  {journal} {\bibinfo
   {journal} {Phys. Rev. E}\ }\textbf {\bibinfo {volume} {105}},\ \bibinfo
  {pages} {014315} (\bibinfo {year} {2022})}\BibitemShut {NoStop}%
\bibitem [{\citenamefont {Kishore}\ \emph {et~al.}(2012)\citenamefont
  {Kishore}, \citenamefont {Santhanam},\ and\ \citenamefont
  {Amritkar}}]{KishoreSanthanamAmritkar2012}%
  \BibitemOpen
  \bibfield  {author} {\bibinfo {author} {\bibfnamefont {V.}~\bibnamefont
  {Kishore}}, \bibinfo {author} {\bibfnamefont {M.~S.}\ \bibnamefont
  {Santhanam}},\ and\ \bibinfo {author} {\bibfnamefont {R.~E.}\ \bibnamefont
  {Amritkar}},\ }\bibfield  {title} {\bibinfo {title} {Extreme events and event
  size fluctuations in biased random walks on networks},\ }\href
  {https://doi.org/10.1103/PhysRevE.85.056120} {\bibfield  {journal} {\bibinfo
  {journal} {Phys. Rev. E}\ }\textbf {\bibinfo {volume} {85}},\ \bibinfo
  {pages} {056120} (\bibinfo {year} {2012})}\BibitemShut {NoStop}%
\bibitem [{\citenamefont {Kumar}\ \emph {et~al.}(2020)\citenamefont {Kumar},
  \citenamefont {Kulkarni},\ and\ \citenamefont
  {Santhanam}}]{AanjSanthanamKulkarni2020}%
  \BibitemOpen
  \bibfield  {author} {\bibinfo {author} {\bibfnamefont {A.}~\bibnamefont
  {Kumar}}, \bibinfo {author} {\bibfnamefont {S.}~\bibnamefont {Kulkarni}},\
  and\ \bibinfo {author} {\bibfnamefont {M.~S.}\ \bibnamefont {Santhanam}},\
  }\bibfield  {title} {\bibinfo {title} {Extreme events in stochastic transport
  on networks},\ }\href {https://doi.org/10.1063/1.5139018} {\bibfield
  {journal} {\bibinfo  {journal} {Chaos: An Interdisciplinary Journal of
  Nonlinear Science}\ }\textbf {\bibinfo {volume} {30}},\ \bibinfo {pages}
  {043111} (\bibinfo {year} {2020})}\BibitemShut {NoStop}%
\bibitem [{\citenamefont {Buarque}\ \emph {et~al.}(2022)\citenamefont
  {Buarque}, \citenamefont {Dias}, \citenamefont {de~Moura}, \citenamefont
  {Lyra},\ and\ \citenamefont {Almeida}}]{rougewave_1d}%
  \BibitemOpen
  \bibfield  {author} {\bibinfo {author} {\bibfnamefont {A.~R.~C.}\
  \bibnamefont {Buarque}}, \bibinfo {author} {\bibfnamefont {W.~S.}\
  \bibnamefont {Dias}}, \bibinfo {author} {\bibfnamefont {F.~A. B.~F.}\
  \bibnamefont {de~Moura}}, \bibinfo {author} {\bibfnamefont {M.~L.}\
  \bibnamefont {Lyra}},\ and\ \bibinfo {author} {\bibfnamefont {G.~M.~A.}\
  \bibnamefont {Almeida}},\ }\bibfield  {title} {\bibinfo {title} {Rogue waves
  in discrete-time quantum walks},\ }\href
  {https://doi.org/10.1103/PhysRevA.106.012414} {\bibfield  {journal} {\bibinfo
   {journal} {Phys. Rev. A}\ }\textbf {\bibinfo {volume} {106}},\ \bibinfo
  {pages} {012414} (\bibinfo {year} {2022})}\BibitemShut {NoStop}%
\bibitem [{\citenamefont {Buarque}\ \emph {et~al.}(2023)\citenamefont
  {Buarque}, \citenamefont {Dias}, \citenamefont {Almeida}, \citenamefont
  {Lyra},\ and\ \citenamefont {de~Moura}}]{BuaDiaAlm2023}%
  \BibitemOpen
  \bibfield  {author} {\bibinfo {author} {\bibfnamefont {A.~R.~C.}\
  \bibnamefont {Buarque}}, \bibinfo {author} {\bibfnamefont {W.~S.}\
  \bibnamefont {Dias}}, \bibinfo {author} {\bibfnamefont {G.~M.~A.}\
  \bibnamefont {Almeida}}, \bibinfo {author} {\bibfnamefont {M.~L.}\
  \bibnamefont {Lyra}},\ and\ \bibinfo {author} {\bibfnamefont {F.~A. B.~F.}\
  \bibnamefont {de~Moura}},\ }\bibfield  {title} {\bibinfo {title} {Rogue waves
  in quantum lattices with correlated disorder},\ }\href
  {https://doi.org/10.1103/PhysRevA.107.012425} {\bibfield  {journal} {\bibinfo
   {journal} {Phys. Rev. A}\ }\textbf {\bibinfo {volume} {107}},\ \bibinfo
  {pages} {012425} (\bibinfo {year} {2023})}\BibitemShut {NoStop}%
\bibitem [{\citenamefont {Buarque}\ and\ \citenamefont
  {Raposo}(2023)}]{BuaRap2023}%
  \BibitemOpen
  \bibfield  {author} {\bibinfo {author} {\bibfnamefont {A.~R.~C.}\
  \bibnamefont {Buarque}}\ and\ \bibinfo {author} {\bibfnamefont {E.~P.}\
  \bibnamefont {Raposo}},\ }\bibfield  {title} {\bibinfo {title} {Managing
  rogue quantum amplitudes: A possible control in quantum walks},\ }\href
  {https://doi.org/10.1103/PhysRevA.108.062206} {\bibfield  {journal} {\bibinfo
   {journal} {Phys. Rev. A}\ }\textbf {\bibinfo {volume} {108}},\ \bibinfo
  {pages} {062206} (\bibinfo {year} {2023})}\BibitemShut {NoStop}%
\bibitem [{\citenamefont {Brun}\ \emph
  {et~al.}(2003{\natexlab{a}})\citenamefont {Brun}, \citenamefont {Carteret},\
  and\ \citenamefont {Ambainis}}]{BrunCartretAmbainis2003}%
  \BibitemOpen
  \bibfield  {author} {\bibinfo {author} {\bibfnamefont {T.~A.}\ \bibnamefont
  {Brun}}, \bibinfo {author} {\bibfnamefont {H.~A.}\ \bibnamefont {Carteret}},\
  and\ \bibinfo {author} {\bibfnamefont {A.}~\bibnamefont {Ambainis}},\
  }\bibfield  {title} {\bibinfo {title} {Quantum walks driven by many coins},\
  }\href {https://doi.org/10.1103/PhysRevA.67.052317} {\bibfield  {journal}
  {\bibinfo  {journal} {Phys. Rev. A}\ }\textbf {\bibinfo {volume} {67}},\
  \bibinfo {pages} {052317} (\bibinfo {year} {2003}{\natexlab{a}})}\BibitemShut
  {NoStop}%
\bibitem [{\citenamefont {Brun}\ \emph
  {et~al.}(2003{\natexlab{b}})\citenamefont {Brun}, \citenamefont {Carteret},\
  and\ \citenamefont {Ambainis}}]{BrunCartretAmbainis2003_QtoCl_transition}%
  \BibitemOpen
  \bibfield  {author} {\bibinfo {author} {\bibfnamefont {T.~A.}\ \bibnamefont
  {Brun}}, \bibinfo {author} {\bibfnamefont {H.~A.}\ \bibnamefont {Carteret}},\
  and\ \bibinfo {author} {\bibfnamefont {A.}~\bibnamefont {Ambainis}},\
  }\bibfield  {title} {\bibinfo {title} {Quantum to classical transition for
  random walks},\ }\href {https://doi.org/10.1103/PhysRevLett.91.130602}
  {\bibfield  {journal} {\bibinfo  {journal} {Phys. Rev. Lett.}\ }\textbf
  {\bibinfo {volume} {91}},\ \bibinfo {pages} {130602} (\bibinfo {year}
  {2003}{\natexlab{b}})}\BibitemShut {NoStop}%
\bibitem [{\citenamefont {Omanakuttan}\ and\ \citenamefont
  {Lakshminarayan}(2021)}]{Arul2021_quantumChaoticCoins}%
  \BibitemOpen
  \bibfield  {author} {\bibinfo {author} {\bibfnamefont {S.}~\bibnamefont
  {Omanakuttan}}\ and\ \bibinfo {author} {\bibfnamefont {A.}~\bibnamefont
  {Lakshminarayan}},\ }\bibfield  {title} {\bibinfo {title} {Quantum walks with
  quantum chaotic coins: Loschmidt echo, classical limit, and thermalization},\
  }\href {https://doi.org/10.1103/PhysRevE.103.012207} {\bibfield  {journal}
  {\bibinfo  {journal} {Phys. Rev. E}\ }\textbf {\bibinfo {volume} {103}},\
  \bibinfo {pages} {012207} (\bibinfo {year} {2021})}\BibitemShut {NoStop}%
\bibitem [{\citenamefont {Kendon}(2007)}]{Kendon2007}%
  \BibitemOpen
  \bibfield  {author} {\bibinfo {author} {\bibfnamefont {V.}~\bibnamefont
  {Kendon}},\ }\bibfield  {title} {\bibinfo {title} {Decoherence in quantum
  walks--a review},\ }\href@noop {} {\bibfield  {journal} {\bibinfo  {journal}
  {Mathematical structures in computer science}\ }\textbf {\bibinfo {volume}
  {17}},\ \bibinfo {pages} {1169} (\bibinfo {year} {2007})}\BibitemShut
  {NoStop}%
\bibitem [{\citenamefont {Bludov}\ \emph {et~al.}(2009)\citenamefont {Bludov},
  \citenamefont {Konotop},\ and\ \citenamefont {Akhmediev}}]{BluKonAkh2009}%
  \BibitemOpen
  \bibfield  {author} {\bibinfo {author} {\bibfnamefont {Y.~V.}\ \bibnamefont
  {Bludov}}, \bibinfo {author} {\bibfnamefont {V.~V.}\ \bibnamefont
  {Konotop}},\ and\ \bibinfo {author} {\bibfnamefont {N.}~\bibnamefont
  {Akhmediev}},\ }\bibfield  {title} {\bibinfo {title} {Matter rogue waves},\
  }\href {https://doi.org/10.1103/PhysRevA.80.033610} {\bibfield  {journal}
  {\bibinfo  {journal} {Phys. Rev. A}\ }\textbf {\bibinfo {volume} {80}},\
  \bibinfo {pages} {033610} (\bibinfo {year} {2009})}\BibitemShut {NoStop}%
\bibitem [{\citenamefont {Yan}\ \emph {et~al.}(2010)\citenamefont {Yan},
  \citenamefont {Konotop},\ and\ \citenamefont {Akhmediev}}]{YanKonAkh2010}%
  \BibitemOpen
  \bibfield  {author} {\bibinfo {author} {\bibfnamefont {Z.}~\bibnamefont
  {Yan}}, \bibinfo {author} {\bibfnamefont {V.~V.}\ \bibnamefont {Konotop}},\
  and\ \bibinfo {author} {\bibfnamefont {N.}~\bibnamefont {Akhmediev}},\
  }\bibfield  {title} {\bibinfo {title} {Three-dimensional rogue waves in
  nonstationary parabolic potentials},\ }\href
  {https://doi.org/10.1103/PhysRevE.82.036610} {\bibfield  {journal} {\bibinfo
  {journal} {Phys. Rev. E}\ }\textbf {\bibinfo {volume} {82}},\ \bibinfo
  {pages} {036610} (\bibinfo {year} {2010})}\BibitemShut {NoStop}%
\bibitem [{\citenamefont {Dematteis}\ \emph {et~al.}(2019)\citenamefont
  {Dematteis}, \citenamefont {Grafke}, \citenamefont {Onorato},\ and\
  \citenamefont {Vanden-Eijnden}}]{DemGraOno2019}%
  \BibitemOpen
  \bibfield  {author} {\bibinfo {author} {\bibfnamefont {G.}~\bibnamefont
  {Dematteis}}, \bibinfo {author} {\bibfnamefont {T.}~\bibnamefont {Grafke}},
  \bibinfo {author} {\bibfnamefont {M.}~\bibnamefont {Onorato}},\ and\ \bibinfo
  {author} {\bibfnamefont {E.}~\bibnamefont {Vanden-Eijnden}},\ }\bibfield
  {title} {\bibinfo {title} {Experimental evidence of hydrodynamic instantons:
  The universal route to rogue waves},\ }\href
  {https://doi.org/10.1103/PhysRevX.9.041057} {\bibfield  {journal} {\bibinfo
  {journal} {Phys. Rev. X}\ }\textbf {\bibinfo {volume} {9}},\ \bibinfo {pages}
  {041057} (\bibinfo {year} {2019})}\BibitemShut {NoStop}%
\bibitem [{\citenamefont {Ding}\ \emph {et~al.}(2024)\citenamefont {Ding},
  \citenamefont {Zhou},\ and\ \citenamefont {Malomed}}]{DinZhoMal2024}%
  \BibitemOpen
  \bibfield  {author} {\bibinfo {author} {\bibfnamefont {C.-C.}\ \bibnamefont
  {Ding}}, \bibinfo {author} {\bibfnamefont {Q.}~\bibnamefont {Zhou}},\ and\
  \bibinfo {author} {\bibfnamefont {B.~A.}\ \bibnamefont {Malomed}},\
  }\bibfield  {title} {\bibinfo {title} {Ultra-high-amplitude peregrine
  solitons induced by helicoidal spin-orbit coupling},\ }\href
  {https://doi.org/10.1103/PhysRevResearch.6.L032036} {\bibfield  {journal}
  {\bibinfo  {journal} {Phys. Rev. Res.}\ }\textbf {\bibinfo {volume} {6}},\
  \bibinfo {pages} {L032036} (\bibinfo {year} {2024})}\BibitemShut {NoStop}%
\bibitem [{\citenamefont {Kundu}\ \emph {et~al.}(2022)\citenamefont {Kundu},
  \citenamefont {Ghosh},\ and\ \citenamefont {Roy}}]{KunGhoRoy2022}%
  \BibitemOpen
  \bibfield  {author} {\bibinfo {author} {\bibfnamefont {N.}~\bibnamefont
  {Kundu}}, \bibinfo {author} {\bibfnamefont {S.}~\bibnamefont {Ghosh}},\ and\
  \bibinfo {author} {\bibfnamefont {U.}~\bibnamefont {Roy}},\ }\bibfield
  {title} {\bibinfo {title} {Quantum simulation of rogue waves in bose-einstein
  condensate: An exact analytical method},\ }\href
  {https://doi.org/https://doi.org/10.1016/j.physleta.2022.128335} {\bibfield
  {journal} {\bibinfo  {journal} {Physics Letters A}\ }\textbf {\bibinfo
  {volume} {449}},\ \bibinfo {pages} {128335} (\bibinfo {year}
  {2022})}\BibitemShut {NoStop}%
\bibitem [{\citenamefont {Izrailev}\ and\ \citenamefont
  {Krokhin}(1999)}]{IzrKro1999}%
  \BibitemOpen
  \bibfield  {author} {\bibinfo {author} {\bibfnamefont {F.~M.}\ \bibnamefont
  {Izrailev}}\ and\ \bibinfo {author} {\bibfnamefont {A.~A.}\ \bibnamefont
  {Krokhin}},\ }\bibfield  {title} {\bibinfo {title} {Localization and the
  mobility edge in one-dimensional potentials with correlated disorder},\
  }\href {https://doi.org/10.1103/PhysRevLett.82.4062} {\bibfield  {journal}
  {\bibinfo  {journal} {Phys. Rev. Lett.}\ }\textbf {\bibinfo {volume} {82}},\
  \bibinfo {pages} {4062} (\bibinfo {year} {1999})}\BibitemShut {NoStop}%
\bibitem [{\citenamefont {Kishore}\ \emph {et~al.}(2011)\citenamefont
  {Kishore}, \citenamefont {Santhanam},\ and\ \citenamefont
  {Amritkar}}]{santhanam}%
  \BibitemOpen
  \bibfield  {author} {\bibinfo {author} {\bibfnamefont {V.}~\bibnamefont
  {Kishore}}, \bibinfo {author} {\bibfnamefont {M.~S.}\ \bibnamefont
  {Santhanam}},\ and\ \bibinfo {author} {\bibfnamefont {R.~E.}\ \bibnamefont
  {Amritkar}},\ }\bibfield  {title} {\bibinfo {title} {Extreme events on
  complex networks},\ }\href {https://doi.org/10.1103/PhysRevLett.106.188701}
  {\bibfield  {journal} {\bibinfo  {journal} {Phys. Rev. Lett.}\ }\textbf
  {\bibinfo {volume} {106}},\ \bibinfo {pages} {188701} (\bibinfo {year}
  {2011})}\BibitemShut {NoStop}%
\bibitem [{\citenamefont {Bonatto}\ \emph {et~al.}(2011)\citenamefont
  {Bonatto}, \citenamefont {Feyereisen}, \citenamefont {Barland}, \citenamefont
  {Giudici}, \citenamefont {Masoller}, \citenamefont {Leite},\ and\
  \citenamefont {Tredicce}}]{deterministic_nonLin_EE}%
  \BibitemOpen
  \bibfield  {author} {\bibinfo {author} {\bibfnamefont {C.}~\bibnamefont
  {Bonatto}}, \bibinfo {author} {\bibfnamefont {M.}~\bibnamefont {Feyereisen}},
  \bibinfo {author} {\bibfnamefont {S.}~\bibnamefont {Barland}}, \bibinfo
  {author} {\bibfnamefont {M.}~\bibnamefont {Giudici}}, \bibinfo {author}
  {\bibfnamefont {C.}~\bibnamefont {Masoller}}, \bibinfo {author}
  {\bibfnamefont {J.~R.~R.}\ \bibnamefont {Leite}},\ and\ \bibinfo {author}
  {\bibfnamefont {J.~R.}\ \bibnamefont {Tredicce}},\ }\bibfield  {title}
  {\bibinfo {title} {Deterministic optical rogue waves},\ }\href
  {https://doi.org/10.1103/PhysRevLett.107.053901} {\bibfield  {journal}
  {\bibinfo  {journal} {Phys. Rev. Lett.}\ }\textbf {\bibinfo {volume} {107}},\
  \bibinfo {pages} {053901} (\bibinfo {year} {2011})}\BibitemShut {NoStop}%
\bibitem [{\citenamefont {Arecchi}\ \emph {et~al.}(2011)\citenamefont
  {Arecchi}, \citenamefont {Bortolozzo}, \citenamefont {Montina},\ and\
  \citenamefont {Residori}}]{EE_inhomogeneity}%
  \BibitemOpen
  \bibfield  {author} {\bibinfo {author} {\bibfnamefont {F.~T.}\ \bibnamefont
  {Arecchi}}, \bibinfo {author} {\bibfnamefont {U.}~\bibnamefont {Bortolozzo}},
  \bibinfo {author} {\bibfnamefont {A.}~\bibnamefont {Montina}},\ and\ \bibinfo
  {author} {\bibfnamefont {S.}~\bibnamefont {Residori}},\ }\bibfield  {title}
  {\bibinfo {title} {Granularity and inhomogeneity are the joint generators of
  optical rogue waves},\ }\href
  {https://doi.org/10.1103/PhysRevLett.106.153901} {\bibfield  {journal}
  {\bibinfo  {journal} {Phys. Rev. Lett.}\ }\textbf {\bibinfo {volume} {106}},\
  \bibinfo {pages} {153901} (\bibinfo {year} {2011})}\BibitemShut {NoStop}%
\bibitem [{\citenamefont {Aharonov}\ \emph {et~al.}(2001)\citenamefont
  {Aharonov}, \citenamefont {Ambainis}, \citenamefont {Kempe},\ and\
  \citenamefont {Vazirani}}]{aharonov}%
  \BibitemOpen
  \bibfield  {author} {\bibinfo {author} {\bibfnamefont {D.}~\bibnamefont
  {Aharonov}}, \bibinfo {author} {\bibfnamefont {A.}~\bibnamefont {Ambainis}},
  \bibinfo {author} {\bibfnamefont {J.}~\bibnamefont {Kempe}},\ and\ \bibinfo
  {author} {\bibfnamefont {U.}~\bibnamefont {Vazirani}},\ }\bibfield  {title}
  {\bibinfo {title} {Quantum walks on graphs},\ }in\ \href
  {https://doi.org/10.1145/380752.380758} {\emph {\bibinfo {booktitle}
  {Proceedings of the Thirty-Third Annual ACM Symposium on Theory of
  Computing}}},\ \bibinfo {series and number} {STOC '01}\ (\bibinfo
  {publisher} {Association for Computing Machinery},\ \bibinfo {address} {New
  York, NY, USA},\ \bibinfo {year} {2001})\ p.\ \bibinfo {pages}
  {50–59}\BibitemShut {NoStop}%
\bibitem [{\citenamefont {Godsil}\ and\ \citenamefont {Zhan}(2019)}]{godsil}%
  \BibitemOpen
  \bibfield  {author} {\bibinfo {author} {\bibfnamefont {C.}~\bibnamefont
  {Godsil}}\ and\ \bibinfo {author} {\bibfnamefont {H.}~\bibnamefont {Zhan}},\
  }\bibfield  {title} {\bibinfo {title} {Discrete-time quantum walks and graph
  structures},\ }\href
  {https://doi.org/https://doi.org/10.1016/j.jcta.2019.05.003} {\bibfield
  {journal} {\bibinfo  {journal} {Journal of Combinatorial Theory, Series A}\
  }\textbf {\bibinfo {volume} {167}},\ \bibinfo {pages} {181} (\bibinfo {year}
  {2019})}\BibitemShut {NoStop}%
\bibitem [{sup()}]{supp_material}%
  \BibitemOpen
  \href@noop {} {}\bibinfo {note} {See Supplemental Material at
  URL-will-be-inserted-by-publisher for the calculations of the
  paper.}\BibitemShut {Stop}%
\bibitem [{\citenamefont {Koll\'ar}\ \emph {et~al.}(2020)\citenamefont
  {Koll\'ar}, \citenamefont {Gily\'en}, \citenamefont
  {Tk\'a\ifmmode~\check{c}\else \v{c}\fi{}ov\'a}, \citenamefont {Kiss},
  \citenamefont {Jex},\ and\ \citenamefont {\ifmmode \check{S}\else
  \v{S}\fi{}tefa\ifmmode~\check{n}\else
  \v{n}\fi{}\'ak}}]{KollarGilyKissStef2020}%
  \BibitemOpen
  \bibfield  {author} {\bibinfo {author} {\bibfnamefont {B.}~\bibnamefont
  {Koll\'ar}}, \bibinfo {author} {\bibfnamefont {A.}~\bibnamefont {Gily\'en}},
  \bibinfo {author} {\bibfnamefont {I.}~\bibnamefont
  {Tk\'a\ifmmode~\check{c}\else \v{c}\fi{}ov\'a}}, \bibinfo {author}
  {\bibfnamefont {T.}~\bibnamefont {Kiss}}, \bibinfo {author} {\bibfnamefont
  {I.}~\bibnamefont {Jex}},\ and\ \bibinfo {author} {\bibfnamefont
  {M.}~\bibnamefont {\ifmmode \check{S}\else
  \v{S}\fi{}tefa\ifmmode~\check{n}\else \v{n}\fi{}\'ak}},\ }\bibfield  {title}
  {\bibinfo {title} {Complete classification of trapping coins for quantum
  walks on the two-dimensional square lattice},\ }\href
  {https://doi.org/10.1103/PhysRevA.102.012207} {\bibfield  {journal} {\bibinfo
   {journal} {Phys. Rev. A}\ }\textbf {\bibinfo {volume} {102}},\ \bibinfo
  {pages} {012207} (\bibinfo {year} {2020})}\BibitemShut {NoStop}%
\bibitem [{\citenamefont {Inui}\ \emph {et~al.}(2004)\citenamefont {Inui},
  \citenamefont {Konishi},\ and\ \citenamefont {Konno}}]{InuiKonishiKonno2004}%
  \BibitemOpen
  \bibfield  {author} {\bibinfo {author} {\bibfnamefont {N.}~\bibnamefont
  {Inui}}, \bibinfo {author} {\bibfnamefont {Y.}~\bibnamefont {Konishi}},\ and\
  \bibinfo {author} {\bibfnamefont {N.}~\bibnamefont {Konno}},\ }\bibfield
  {title} {\bibinfo {title} {Localization of two-dimensional quantum walks},\
  }\href {https://doi.org/10.1103/PhysRevA.69.052323} {\bibfield  {journal}
  {\bibinfo  {journal} {Phys. Rev. A}\ }\textbf {\bibinfo {volume} {69}},\
  \bibinfo {pages} {052323} (\bibinfo {year} {2004})}\BibitemShut {NoStop}%
\end{thebibliography}%
\balancecolsandclearpage
\clearpage
\onecolumngrid
\begin{center}
    \textbf{\Huge{Supplementary Material I}}
\end{center}

\vspace{2em}
To perform a DTQW on an arbitrary graph $G(V,E)$ with a vertex set $V$ and an edge set $E$, the undirected graph is viewed as a directed one and the quantum walk takes place in a complex vector space $\mathbf{C}^{2\vert E \vert}$ which is spanned by the basis $e_{ij}$ representing the edge between vertices $i$ and $j$. The coin at each vertex, $u$ acts to replace the amplitude on each edge with a superposition of that on all the other edges of $u$. For a graph with $N$ vertices, $\vert V \vert  = N$, and the coin operator $C$ is
\begin{equation*}
C = diag\{ C_1, C_2, ..., C_N \}    ,
\end{equation*}
where each $C_i$ is a unitary matrix of order $k_i$, the degree of $i$-th vertex.

\section{Operators}
\paragraph{Coin operators}
The Fourier coin is a generalization of the Hadamard coin to arbitrary size, and is defined as:
$$
C_i^{\mathrm{F}}\left(\begin{array}{c}
\left|i \rightarrow j_1\right\rangle \\
\left|i \rightarrow j_2\right\rangle \\
\left|i \rightarrow j_3\right\rangle \\
\vdots \\
\left|i \rightarrow j_{k_i}\right\rangle
\end{array}\right)=\frac{1}{\sqrt{k_i}}\left(\begin{array}{ccccc}
1 & 1 & 1 & \cdots & 1 \\
1 & e^{i \theta / k_i} & e^{2 i \theta / k_i} & \cdots & e^{\left(k_i-1\right) i \theta / k_i} \\
1 & e^{2 i \theta / k_i} & e^{4 i \theta / k_i} & \cdots & e^{2\left(k_i-1\right) i \theta / k_i} \\
\vdots & \vdots & \vdots & \ddots & \vdots \\
1 & e^{\left(k_i-1\right) i \theta / k_i} & e^{2\left(k_i-1\right) i \theta / k_i} & \cdots & e^{\left(k_i-1\right)\left(k_i-1\right) i \theta / k_i}
\end{array}\right)\left(\begin{array}{c}
\left|i \rightarrow j_1\right\rangle \\
\left|i \rightarrow j_2\right\rangle \\
\left|i \rightarrow j_3\right\rangle \\
\vdots \\
\left|i \rightarrow j_{k_i}\right\rangle
\end{array}\right)
$$
Grover coin is defined as:
$$
C_i^{\mathrm{G}}\left(\begin{array}{c}
\left|i \rightarrow j_1\right\rangle \\
\left|i \rightarrow j_2\right\rangle \\
\left|i \rightarrow j_3\right\rangle \\
\vdots \\
\left|i \rightarrow j_{k_i}\right\rangle
\end{array}\right)=\frac{1}{k_i}\left(\begin{array}{ccccc}
2-k_i & 2 & 2 & 2 & 2 \\
2 & 2-k_i & 2 & 2 & 2 \\
2 & 2 & 2-k_i & 2 & 2 \\
\vdots & \vdots & \vdots & \ddots & \vdots \\
2 & 2 & 2 & \cdots & 2-k_i
\end{array}\right)\left(\begin{array}{c}
\left|i \rightarrow j_1\right\rangle \\
\left|i \rightarrow j_2\right\rangle \\
\left|i \rightarrow j_3\right\rangle \\
\vdots \\
\left|i \rightarrow j_{k_i}\right\rangle
\end{array}\right)
$$
where the $j_l \ \ l \in \{1,2,...k_i\}$ are the vertices adjacent to the $i$-th vertex.
\paragraph{Shift operator}
The shift operator used in the present study is such that it swaps the amplitudes on adjacent nodes. It can be defined through its action on the basis as: $$S(e_{ij}) = e_{ji}.$$
For a periodic 1D lattice, each site has two states, left $\lvert -\rangle$, and right $\lvert +\rangle$, owing to the two out-going edges (to the two neighbors either side). At each site, the shift operator acts to swap the amplitude of the $\lvert +\rangle$ state with that of the right neighbor, and the amplitude of the $\lvert -\rangle$ state with that of the left neighbor. In the computational basis of the Hilbert space $\mathcal{H}_p \otimes \mathcal{H}_c$, the alternate entries of a state vector $\lvert x\rangle$ represent the $\lvert -\rangle$ and $\lvert +\rangle$ amplitudes at a position. The following matrix shows an example shift operator for a periodic 1D lattice of size 3. The first row can be read off as mapping the $\vert-\rangle$ state of $3^{rd}$ position to that of the $1^{st}$.
\begin{equation*}
\begin{bmatrix}
0 & 0 & 0 & 0 & 1 & 0 \\
0 & 0 & 0 & 1 & 0 & 0 \\
1 & 0 & 0 & 0 & 0 & 0 \\
0 & 0 & 0 & 0 & 0 & 1 \\
0 & 0 & 1 & 0 & 0 & 0 \\
0 & 1 & 0 & 0 & 0 & 0 \\
\end{bmatrix}
\end{equation*}
\section{Theorem on Average Limiting Distribution}
Define the probability amplitude on a vertex $v$ as the sum of the probabilities on all of the edges incident on $v$. Then this can be written as a projection of the evolved state vector onto the subspace spaned by $\{ e_{v,1}\ e_{v,2}\ \cdots \ e_{v,deg(v)} \}$. 
$$
\left(\begin{array}{llll}
e_{v,1} & e_{v,2} & \cdots & e_{v,{deg(v)}}
\end{array}\right)^{\top} U^t \lvert x \rangle
$$
Define:
$$
D_v=\left(\begin{array}{llll}
e_{v,1} & e_{v,2} & \cdots & e_{v,deg(v)}
\end{array}\right)\left(\begin{array}{llll}
e_{v,1} & e_{v,2} & \cdots & e_{v,deg(v)}
\end{array}\right)^{\top}
$$
That is, $D_v$ is diagonal matrix whose $(i,i)$ entry is $1$ if the $i^{th}$ entry of the state vector represents some outgoing edge of $v$. Given an initial state $\lvert x \rangle$, the probability amplitude at vertex $v$ at time $t$ is given as:
$$
P_{x,D_v}(t) = \langle x \rvert \left(U^t\right)^* D_v U^t \lvert x \rangle.
$$
Omitting the reference to the initial state for the rest of the material we adapt the following notation:
$$z_{v}(t) = \langle x \rvert \left(U^t\right)^* D_v U^t \lvert x \rangle.$$
\textbf{Theorem}\\
Let $F_1, F_2, \ldots, F_m$ be the spectral idempotents of the transition matrix $U$. Let $\lvert x\rangle$ be the initial state, then for any vertex $v$, the time average probability at the node $v$ converges to $\sum_r \langle x \rvert F_r D_v F_r \lvert x \rangle $. That is, 
\begin{equation}\label{eq:convergence}
    \lim_{T\to\infty} \frac{1}{T} \sum_{t=0}^{T-1}z_{v}(t) = \sum_r \langle x \rvert F_r D_v F_r \lvert x \rangle
\end{equation}
\textbf{Proof \cite{aharonov}}\\
Every unitary matrix, $U$ can be written as $U=\sum_{r} e^{i\theta_{r}} F_r$ where $e^{i\theta_r}$ is an eigenvalue of $U$, and $F_r$ is a Hermitian matrix representing the projection onto the eigen-space of $e^{i \theta}$. This is usually called the spectral decomposition of $U$. Consider
$$
\begin{aligned}
\left(U^t\right)^* D_v U^t & =\left(\sum_r e^{-i t \theta_r} F_r\right) D_v\left(\sum_s e^{i t \theta_s} F_r\right) \\
& =\sum_r F_r D_S F_r+\sum_{r \neq s} e^{i t\left(\theta_s-\theta_r\right)} F_r D_v F_s .
\end{aligned}
$$
Note that for all $r$ and $s$, the entries in $F_r D_S F_r$ and $F_r D_S F_s$ are constants, and remain unchanged when we take the average and the limit. Further
$$
\frac{1}{T}\left|\sum_{t=0}^{T-1} e^{i\left(\theta_s-\theta_r\right)}\right|=\frac{1}{T}\left|\frac{1-e^{i T\left(\theta_s-\theta_r\right)}}{1-e^{i\left(\theta_s-\theta_r\right)}}\right| \leq \frac{1}{T} \frac{2}{\left|1-e^{i\left(\theta_s-\theta_r\right)}\right|}
$$
if the eigenvalues are non-degenerate, the above term 
converges to zero as $T$ goes to infinity. Hence the only term that survives is $\sum_{r} F_r D_v F_r$. 

\section{Mean of time averaged probability amplitudes at nodes of same degree}
In order to use the above theorem, we inspect the terms in equation (\ref{eq:convergence}) in full generality. Define, $D_v^i$ as a diagonal matrix of order $(m \times m)$, with $1$ at the entry corresponding to the $i^{th}$ edge of a node $v$. Here, we shall use the notation $2\vert E\vert = M$ for ease of notation. Clearly,
$$D_v = \sum_{i = 1}^{deg(v)} D_v^{i}$$
This implies, 
$$
F_r D_v^{i} F_r = 
\left[\begin{array}{cccc}
F_{1i}^rF_{i1}^r & F_{1i}^rF_{i2}^r & \cdots & F_{1i}^rF_{iM}^r \\
F_{2i}^rF_{i1}^r & F_{2i}^rF_{i2}^r & \cdots & F_{2i}^rF_{iM}^r \\
\vdots & \vdots & \ddots& \vdots \\
F_{Mi}^rF_{i1}^r & F_{Mi}^nF_{i2}^r & \cdots & F_{Mi}^rF_{iM}^r
\end{array}\right]
$$
where we change the subscript on $F_r$ to superscript for notational convenience. Since $F_r$ are eigen-space projection operators, in computational basis they can be written as
$$F_r = Q P_r Q^{\dagger}$$
where $P_r$ has all entries zero except at $(r,r)$ and, $Q$ is the unitary matrix that transforms the eigenbasis to computational basis: $Q=\left[ \hat{v}_1, \hat{v}_2, \ldots, \hat{v}_M \right]$, where $\hat{v_i}$ is the $i^{th}$ eigenvector of the unitary transition matrix $U$. 
This allows one to write $F_r$ in terms of components of eigenvectors of $U$ as $(F_r)_{ij} = v_{ir}v^*_{jr}$.
In the next two subsections we calculate the mean for two drastically different initial states, namely, uniformly superposed initial state, and a localized initial state. 
\subsection{Uniform superposition as the initial state.}
Let $\lvert x \rangle = \frac{1}{\sqrt{M}} (1 1 1 ... 1)^{\top}$ be the normalized initial state of the system. Then the time averaged probability amplitude on the $i^{th}$ edge of vertex $v$ of degree $k$ is:
$$\lim_{T\to\infty}\frac{1}{T}\sum_{t=0}^{T-1} z_{v^i}(t) = \sum_{r}\langle x \rvert  \left( F_r D_{v}^{i} F_r \right) \lvert x \rangle  = \frac{1}{M}\sum_{r}\sum_{k,j=1}^{M} F^r_{ki}F^r_{ij}$$
Upon some manipulations, this can be rewritten as
\begin{equation*}
	\sum_r \langle x \rvert \left(F_r D_{v}^i F_r\right) \lvert x \rangle =\frac{1}{M} + \frac{1}{M} \sum_{r}\sum_{k \neq j}^{M} F^r_{ki}F^r_{ij}
\end{equation*}
where we used the orthonormality of eigenvectors of $U$ and that $\sum_r F_r = I$. Now, summing over all the edges of the vertex of degree $k$ gives:
\begin{equation*}
	\sum_r \langle x \rvert \left(F_r D_{v} F_r\right) \lvert x \rangle =\frac{k}{M} + \frac{1}{M} \sum_{i=1}^k\sum_{r}\sum_{k \neq j}^{M} F^r_{ki}F^r_{ij}
\end{equation*} 
Define $S_k = \{v\in V \vert deg(v) = k\}$. Then the probability amplitudes averaged over vertices of same degree can be written as:
$$ 
\frac{1}{\vert S_k\vert}\sum_{\substack{r \\ v\in S_k}} \langle x \rvert \left(F_r D_{v} F_r\right) \lvert x \rangle = \frac{k}{M} + \frac{1}{\vert S_k \vert M}\sum_{\substack{i=1 \\ v\in S_k}}^{k}\sum_r\left( \Vert v_{ir}\Vert^2 \sum_{k\neq j} F^r_{kj} \right)
$$
where we have summed over the edges of $v\in S_k$, and used $(F_r)_{ij} = v_{ir}v^*_{jr}$ for the second term. Note that the first term is the one expected for classical random walks on a graph. The second term here constitutes a deviation and can be shown to be approximately zero. If we assume that the sum $\sum_{i, v\in S_k}^{k} \Vert v_{ir} \Vert^2$ is nearly independent of $r$, then the second term looks like:
$$\frac{1}{\vert S_k \vert M} \sum_{\substack{i=1 \\ v\in S_k}}^{k} \Vert v_{ir}\Vert^2  \left( \sum_r \sum_{k\neq j} F^r_{kj} \right)$$
where the summation in the bracket is zero owing to $\sum_r F_r = I$. 

\subsection{A localized initial state}
To emphasise that the above result is not dependent on the initial state, but only on the assumption about the $\sum_{i, v\in S_k}^{k} \Vert v_{ir} \Vert$ term, consider an initial state which is localised at some node $u$ of degree $d$. The initial state vector shall look like:
\[
\lvert x \rangle = 
			\begin{cases}
			1, & \text{where index corresponds to an edge of } u \\ 
			0, & otherwise
			\end{cases}
\]
Note that the node at which we observe the probability amplitude time-series is still of degree $k$. One can reproduce the calculation to obtain:
\begin{equation*}
    \frac{1}{\vert S_k\vert}\sum_{v\in S_k}\sum_{r} \langle x \rvert \left(F_r D_{v} F_r\right) \lvert x \rangle = \frac{k}{M} \left( 1 + \frac{1}{d}\sum_{r} \sum_{k\neq j}^{d}F^r_{kj}\right)
\end{equation*}
where we used the assumption that $\Vert v_{ir} \Vert^2 \approx 1/M$ on average. Note that the second term in the bracket is zero as for any $k\neq j,\ \ \sum_r F^r_{kj} = 0$. Hence, regardless of the initial state, we have a term which corresponds to the classical random walk result, and a deviation that depends on the choice of initial state and the quantum walk unitary. Hence, under the assumptions made above, the relation between time-averaged probability at a node of given degree $k$ can be summarized as:
\begin{equation}\label{eq:mean_theory}
    \langle z_v \rangle \sim \frac{k}{M}
\end{equation}

\section{Standard deviation of the probability amplitudes at a node}
Variance of a discrete random variable $X(x)$ is defined as: $\sigma^2(X) = \frac{1}{N}\sum_{i=1}^{N}\left( X(x_i) - \langle X \rangle \right)^2$. For the case at hand we can write:
\begin{align*}
\sigma^{2}\left(z_{v}(t)\right) = & \frac{1}{T+1}\sum_{t=0}^{T} \left( z_{v}(t) -  \langle z_{v}\rangle  \right)^2 \\
 = & \frac{1}{T+1}\sum_{t=0}^{T} \sum_{r\neq s}^{M} \left( e^{it(\theta_s - \theta_r)} \langle x \rvert F_r D_v F_s \lvert x \rangle \right)^2
\end{align*}
Assuming initial state to be uniform superposition of all the states and expanding the sum, we get
$$
\sigma^{2} = \frac{1}{T+1} \sum_{0}^{T+1} \left[ \frac{1}{M^2} \sum_{i,i^{\prime} = 1}^{k} \sum_{\substack{r\neq s\\ r^{\prime} \neq s^{\prime}}} \sum_{\substack{k \neq j\\ k^{\prime} \neq j^{\prime}}} e^{it(\tilde{\theta})} F^{r}_{ki}F^{s}_{ij}F^{r^{\prime}}_{k^{\prime}i^{\prime}}F^{s^{\prime}}_{i^{\prime}j^{\prime}}\right]
$$
where $\tilde{\theta} = \theta_s - \theta_r + \theta_{s^{\prime}} - \theta_{r^{\prime}}$. Upon time-averaging, only terms which survive are those for which $\tilde{\theta} \equiv 0 \text{ mod}(2 \pi)$. Since we have assumed non-degeneracy of eigenvalues, this happens when $r = s^{\prime}$ which also implies $s = r^{\prime}$. Thus we have:
$$
\sigma^{2} = \frac{1}{M^2} \sum_{i,i^{\prime} = 1}^{k} \sum_{r\neq s} \sum_{\substack{k \neq j\\ k^{\prime} \neq j^{\prime}}} F^{r}_{ki}F^{s}_{ij}F^{s}_{k^{\prime}i^{\prime}}F^{r}_{i^{\prime}j^{\prime}}
$$
leading contribution when $j=k^{\prime}$:
$$
\begin{aligned}
& \frac{1}{M^2} \sum_{i, i^{\prime}}^k \sum_{r \neq s} \sum_{k, j^{\prime}} F_{k i}^r\left(\sum_j^M F_{i j}^s F_{j i^{\prime}}^s\right) F_{i^{\prime} j^{\prime}}^r \\
& =\frac{1}{M^2} \sum_{i, i^{\prime}}^k \sum_{r \neq s} \sum_{k, j^{\prime}} F_{k i}^r F_{i i^{\prime}}^{s} F_{i^{\prime} j^{\prime}}^r \\
& =\frac{1}{M^2} \sum_{i, i^{\prime}}^k \sum_r \sum_{k, j^{\prime}} F^r_{k i} \underbrace{\sum_{s} F_{i i^{\prime}}^s}_{\delta_{i i^{\prime}}} F_{i^{\prime} j^{\prime}}^r = \frac{1}{M^2} \sum_i^k \sum_r \sum_{k, j^{\prime}} F_{k i}^r F_{i j^{\prime}}^r \\
&
\end{aligned}
$$
recursively using properties of $F_r$, and retaining the leading terms, one gets the dependence of $\sigma$ on the degree $k$ of the node as:
$$
\begin{gathered}
\sigma^2 \sim \frac{1}{M^2} \sum_{i}\sum_r \sum_{k=j^{\prime}} F_{i j}^r F_{j^{\prime} i}^r=\frac{1}{M^2} \sum_{i=1}^k \underbrace{\sum_r^M F_{i i}^r}_{\delta_{i i}} \\
\sigma^2 = \frac{1}{M^2} \sum_{i=1}^k \delta_{i i}=\frac{k}{M^2} \hspace{1em} \Longrightarrow \hspace{1em}
\sigma \sim \frac{\sqrt{k}}{M}
\end{gathered}
$$
Using equation \ref{eq:mean_theory}, one obtains the flux-fluctuation relation for DTQW as:
\begin{equation}\label{eq:sigma}
    \sigma_v \sim \frac{1}{\sqrt{M}} \sqrt{\langle z_v\rangle}
\end{equation}
\section{Lifting the non-degeneracy assumption}
In derivation of the theorem \ref{eq:convergence}, \textit{Aharonov et. al.} \cite{aharonov} made an assumption that the spectra of the unitary evolution operator is non-degenerate. We numerically investigate the effect of lifting that assumption in this section.
The high level of degeneracy in the eigenvalue spectra of the unitary evolution operator with a Grover coin of appropriate order at each site provides a suitable case to study. It has been noted in literature that such a degeneracy leads to partial localization \cite{KollarGilyKissStef2020, InuiKonishiKonno2004}.
\begin{figure}[h]
\label{fig:spread}
\centering
    \includegraphics*[width=0.75\linewidth]{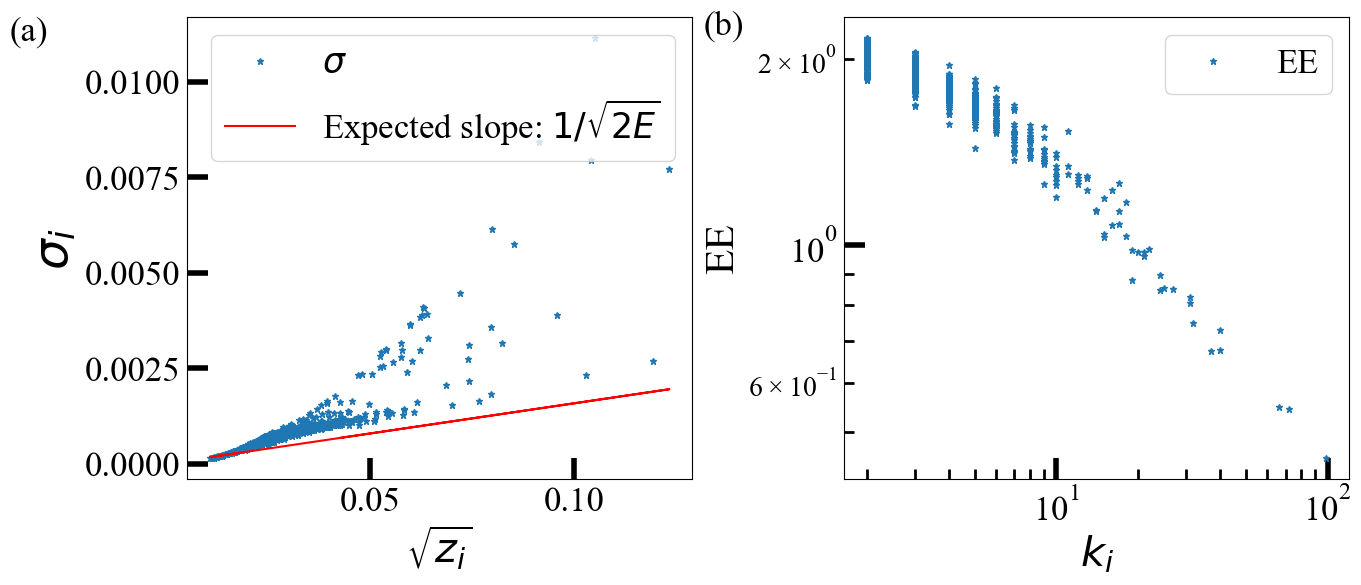}
    \caption{(a) Standard deviation vs. square root of the mean of the time-series of squared probability amplitudes for a DTQW on a SF graph of 1000 nodes. The coin operator used is a direct sum of Grover coins at each vertex. (b)Extreme event probability as a function of vertex degrees on a log-log scale for the same DTQW.}\label{fig:grover}
\end{figure}

\section{Record statistics of extreme events}
How often does EE recur at a node? A systematic treatment of recurrence statistics of EE can provide information about the temporal correlations among EE. An exponentially decaying frequency distribution of recurrence times then implies that EE at various nodes are uncorrelated. 
\begin{figure}[h]
\label{fig:spread}
\centering
    \includegraphics*[width=1\linewidth]{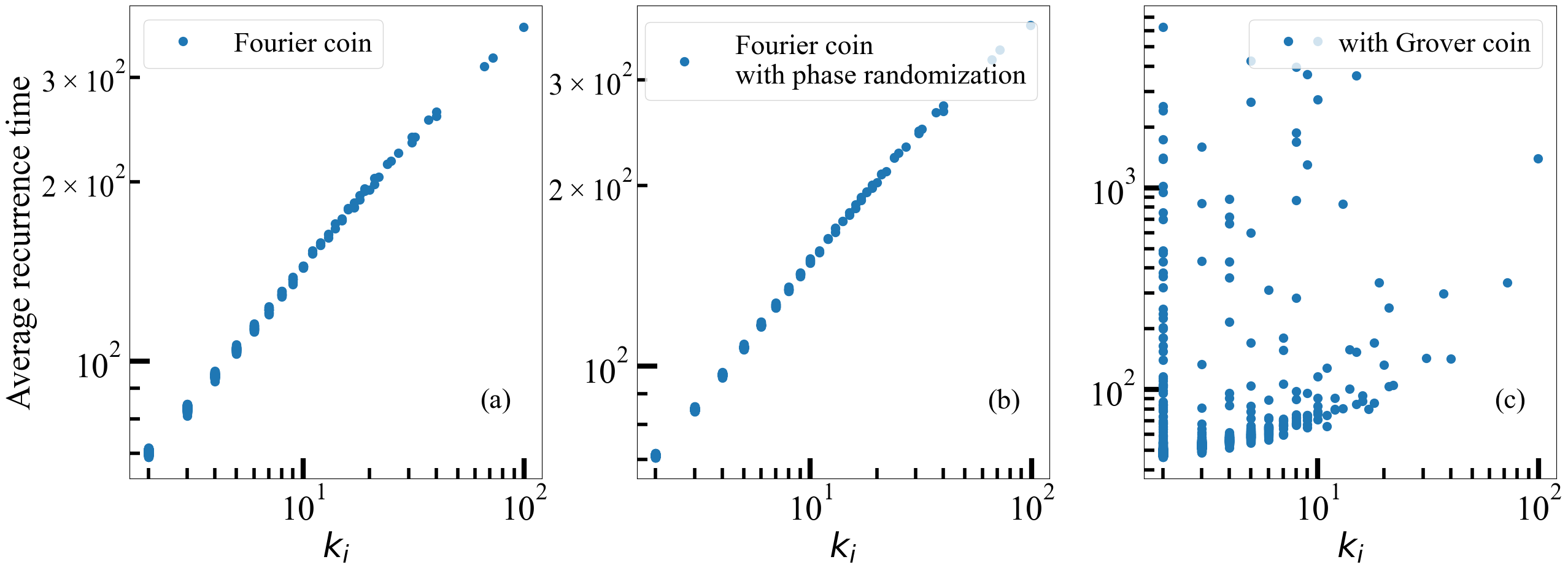}
\caption{Average recurrence time at each node of a scale-free network of 1000 nodes, as a function of the degree of the nodes. The three plots show the results obtained by performing DTQW (a) with Fourier coin (b) with Fourier coin and phase randomization, and (c) with Grover coin.}\label{fig:avg_rec}
\end{figure}

\begin{figure}[h]
\label{fig:spread}
\centering
    \includegraphics*[width=01\linewidth]{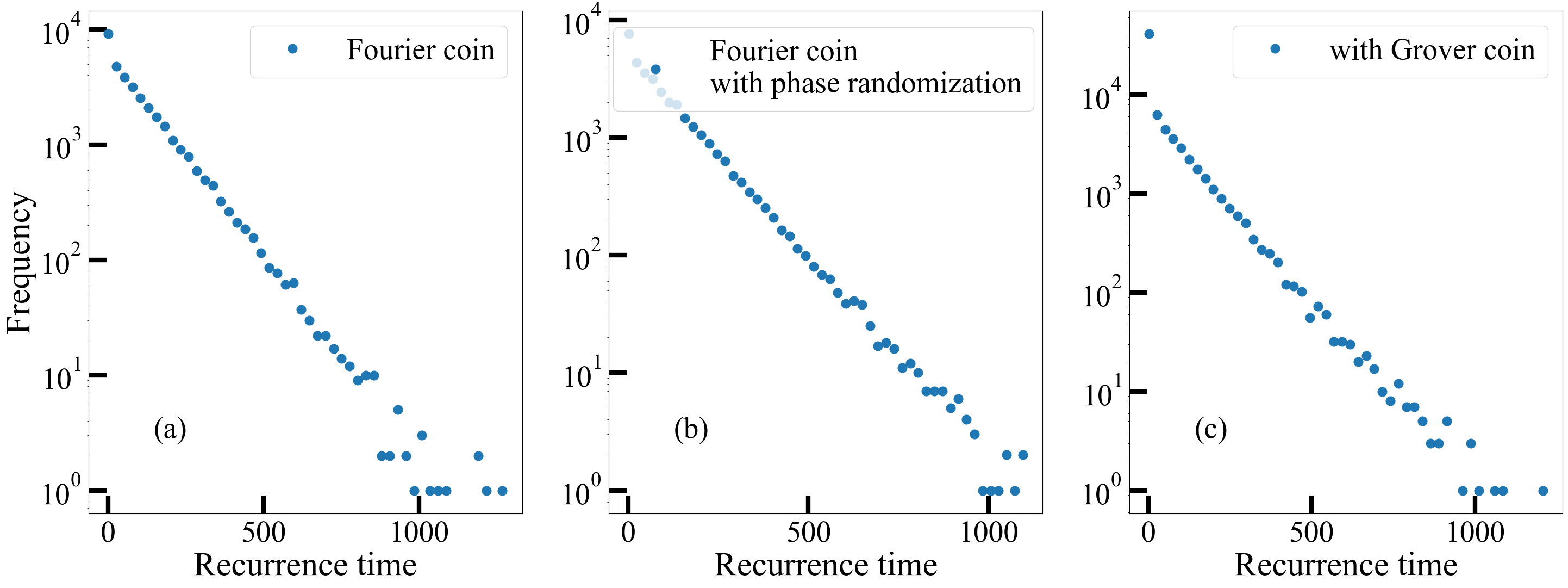}
\caption{Recurrence time distributions at a node of degree 11 in a scale-free network of 1000 nodes. The three plots show the results obtained by performing DTQW (a) with Fourier coin (b) with Fourier coin and phase randomization, and (c) with Grover coin.}\label{fig:rec-time_distribution}
\end{figure}
Note that even though the flux-fluctuation relation (\ref{eq:sigma}) does not hold for the case of DTQW with Grover coin (as evident in Fig. \ref{fig:grover}), the distribution of recurrence times is still a nice exponential as shown in Fig. (\ref{fig:rec-time_distribution}). This is because the null correlation between time-series of squared probability amplitudes is a sufficient condition for obtaining an exponential recurrence-time distribution. However the effect of violation of the flux-fluctuation relation can be captured in Fig. (\ref{fig:avg_rec}) where the dependence of the average recurrence time on the degree of a vertex is affected for the case of DTQW with Grover coin.

\section{PDF of probability amplitudes at a node}
The fraction of time vertex $v$ has probability amplitude $p^{\prime} \in \left[ a, b \right]$:
\begin{equation}\label{eq:invert}
	\int_{a}^{b}\rho(p^{\prime})dp^{\prime} = \frac{1}{T+1} \sum_{t=0}^{T} \int_{a}^{b} \delta \left( p^{\prime}(t) -p^{\prime} \right) d p^{\prime}
\end{equation}
That is to say that if we know $p^{\prime}(t)$ as a function of time alone, we can in principle obtain the PDF of time-series at a node in closed form by inverting the above expression. We elaborate a scheme to calculate $\rho^{\prime}$ in the subsections below.

\subsection{Eigenvalue distribution}
We consider a $(M \times M)$ unitary matrix $U$, and its set of eigenvalues $\lbrace \lambda_r \vert  \lambda_r = e^{i\theta_r}, r \in \lbrace1,2,...,M \rbrace, \theta_r \in [0,2\pi] \rbrace$. 
Let $\sigma(\theta)$ be the density of the eigenvalues around argument $\theta$. The fraction of eigenvalues in the range $[\theta, \theta + d\theta]$ is then: $\sigma(\theta)d\theta$. Number of eigenvalues in the same range of arguments is: $m\times \sigma(\theta)d\theta$.
We need to find the number of eigenvalue pairs as a function of the difference in their arguments. To that effect, we define $\Omega(\omega)$ as the fraction of eigenvalue pairs such that the difference in their arguments lies in the range $[\omega, \omega +d\omega]$. That is:
$$ 
\begin{aligned}
\Omega(\omega)d\omega = & \text{  fraction of eigenvalue pairs  } (\lambda_j, \lambda_k) \\
 			& \text{  such that  } \vert arg(\lambda_j) - arg(\lambda_k)\vert \in [\omega, \omega +d\omega]
\end{aligned}
$$
This quantity can be calculated in terms of $\sigma$ as:
$$
\Omega(\omega) = \iint \sigma(\theta) d \theta \sigma(\varphi) d \varphi \delta(|\varphi-\theta|-\omega)
$$
which reduces to the involution
\begin{equation}\label{eq:omega}
	\Omega(\omega)=2 \int_{\theta=0}^{2 \pi} \sigma(\theta) \sigma(\theta+\omega) d \theta
\end{equation}

\subsection{Calculating $\mathbf{p^{\prime}_{x,v}(t)}$}
Consider the following expression for $p^{\prime}(t)$, where we have initial state as uniform superposition of all states and vertex under consideration has a degree $k$:
$$ 
p^{\prime}_{v}(t) = \frac{1}{M}\sum_{i=1}^{k} \sum_{r\neq s}^{M} e^{it(\theta_s - \theta_r)} \sum_{k\neq j}^{M} F^{r}_{ki}F^{s}_{ij}
$$
Define $\omega_{sr} = \theta_s - \theta_r$ and shuffle the summations to observe that:
$$
p^{\prime}_{v}(t) = \sum_{r\neq s}^{M} e^{it\omega_{sr}} \underbrace{\left( \frac{1}{M}\sum_{i=1}^{k} \sum_{k\neq j}^{M} F^{r}_{ki}F^{s}_{ij} \right) }_{\text{factor f}}
$$
If we assume this factor $f$ to be almost uniform as you vary r and s. Then one can write,
$$
p_v^{\prime}(t) \sim \left(\sum_{r\neq s} e^{i \omega_{sr} }\right) \times \text { f }
$$
Define $p^{\prime\prime}$ as the re-scaled variable:
$$
p^{\prime\prime} := \frac{p^{\prime}}{f}
$$
$$
p(t) = \sum_{r\neq s} e^{i \omega_{sr} } = \sum_{\omega_{sr}\in(0,2 \pi)} \Omega\left(\omega_{sr}\right) e^{i t\left(\omega_{sr}\right)} 
$$
In the limit of large-enough eigenvalue pairs (large unitary matrix), we can approximate the summation with an integral and write:
\begin{equation}\label{eq:fourierlike}
	p(t) = \frac{(M^2-M)}{2} \int_{0}^{2\pi} \Omega (\omega) e^{i t \omega} d \omega
\end{equation}
Equations(\ref{eq:fourierlike}) and (\ref{eq:invert}) together capture the dependence of the PDF on the spectral properties of the evolving unitary via the two-point distribution function $\Omega$ as given in Eq. (\ref{eq:omega}). 

\end{document}